\documentstyle[psfig]{mn}

\newif\ifAMStwofonts

\newcommand{\q}{${\bf q}$}

\newcommand{\pot}{$\varphi$}
\newcommand{\mq}{{\bf q}}

\newcommand{\be}{\begin{equation}}
\newcommand{\ee}{\end{equation}}
\newcommand{\beal}{\begin{equation}\begin{array}{rclcl}}
\newcommand{\eeal}{\end{array}\end{equation}}
\newcommand{\bea}{\begin{eqnarray}}
\newcommand{\eea}{\end{eqnarray}}
\newcommand{\res}{$\Lambda$}
\newcommand{\mres}{\Lambda}
\newcommand{\f}{$F$}
\newcommand{\fc}{$F_c$}
\newcommand{\pf}{$P_F(F,\Lambda)$}
\newcommand{\mpf}{P_F(F,\Lambda)}
\newcommand{\pfu}{$P_F^{\rm noup}(F,\Lambda)$}
\newcommand{\mpfu}{P_F^{\rm noup}(F,\Lambda)}
\newcommand{\ot}{$\Omega(<\Lambda)$}
\newcommand{\mot}{\Omega(<\Lambda)}
\newcommand{\om}{$\Omega(>M)$}

\newcommand{\mf}{$n(M)$}
\newcommand{\mft}{$n(\Lambda)$}
\newcommand{\px}{$P_x(x,\Lambda)$}
\newcommand{\mpx}{P_x(x,\Lambda)}
\newcommand{\pxu}{$P_x^{\rm noup}(x,\Lambda)$}
\newcommand{\mpxu}{P_x^{\rm noup}(x,\Lambda)}
\newcommand{\pxb}{$P_x^{\rm fb}(x,\Lambda)$}
\newcommand{\mpxb}{P_x^{\rm fb}(x,\Lambda)}
\newcommand{\lc}{$\Lambda_c$}

\ifoldfss
  \ifCUPmtlplainloaded \else
    \NewTextAlphabet{textbfit} {cmbxti10} {}
    \NewTextAlphabet{textbfss} {cmssbx10} {}
    \NewMathAlphabet{mathbfit} {cmbxti10} {} 
    \NewMathAlphabet{mathbfss} {cmssbx10} {} 
  \fi
  \ifAMStwofonts
    \ifCUPmtlplainloaded \else
      \NewSymbolFont{upmath} {eurm10}
      \NewSymbolFont{AMSa} {msam10}
      \NewMathSymbol{\upi}     {0}{upmath}{19}
      \NewMathSymbol{\umu}     {0}{upmath}{16}
      \NewMathSymbol{\upartial}{0}{upmath}{40}
      \NewMathSymbol{\leqslant}{3}{AMSa}{36}
      \NewMathSymbol{\geqslant}{3}{AMSa}{3E}

      \let\geq=\geqslant 
    \fi
  \fi
\fi 

\ifnfssone
  \newmathalphabet{\mathit}
  \addtoversion{normal}{\mathit}{cmr}{m}{it}
  \addtoversion{bold}{\mathit}{cmr}{bx}{it}
  \newmathalphabet{\mathbfit} 
  \addtoversion{normal}{\mathbfit}{cmr}{bx}{it}
  \addtoversion{bold}{\mathbfit}{cmr}{bx}{it}
  \newmathalphabet{\mathbfss} 
  \addtoversion{normal}{\mathbfss}{cmss}{bx}{n}
  \addtoversion{bold}{\mathbfss}{cmss}{bx}{n}
  \ifAMStwofonts
    \ifCUPmtlplainloaded \else
      \UseAMStwoboldmath
      \makeatletter
      \new@mathgroup\upmath@group
      \define@mathgroup\mv@normal\upmath@group{eur}{m}{n}
      \define@mathgroup\mv@bold\upmath@group{eur}{b}{n}
      \edef\UPM{\hexnumber\upmath@group}
      \new@mathgroup\amsa@group
      \define@mathgroup\mv@normal\amsa@group{msa}{m}{n}
      \define@mathgroup\mv@bold\amsa@group{msa}{m}{n}
      \edef\AMSa{\hexnumber\amsa@group}
      \makeatother
      \mathchardef\upi="0\UPM19
      \mathchardef\umu="0\UPM16
      \mathchardef\upartial="0\UPM40
      \mathchardef\leqslant="3\AMSa36
      \mathchardef\geqslant="3\AMSa3E

      \let\geq=\geqslant 
    \fi
  \fi
\fi 

\ifnfsstwo
  \DeclareMathAlphabet{\mathbfit}{OT1}{cmr}{bx}{it}
  \SetMathAlphabet\mathbfit{bold}{OT1}{cmr}{bx}{it}
  \DeclareMathAlphabet{\mathbfss}{OT1}{cmss}{bx}{n}
  \SetMathAlphabet\mathbfss{bold}{OT1}{cmss}{bx}{n}
  \ifAMStwofonts
    \ifCUPmtlplainloaded \else
      \DeclareSymbolFont{UPM}{U}{eur}{m}{n}
      \SetSymbolFont{UPM}{bold}{U}{eur}{b}{n}
      \DeclareSymbolFont{AMSa}{U}{msa}{m}{n}
      \DeclareMathSymbol{\upi}{0}{UPM}{"19}
      \DeclareMathSymbol{\umu}{0}{UPM}{"16}
      \DeclareMathSymbol{\upartial}{0}{UPM}{"40}
      \DeclareMathSymbol{\leqslant}{3}{AMSa}{"36}
      \DeclareMathSymbol{\geqslant}{3}{AMSa}{"3E}

      \let\geq=\geqslant 
    \fi
  \fi
\fi 

\ifCUPmtlplainloaded \else
  \ifAMStwofonts \else 
    \def\upi{\pi}
    \def\umu{\mu}
    \def\upartial{\partial}
  \fi
\fi

\title[Mass Function Theory: II Statistics]{A Lagrangian Dynamical Theory 
       for the Mass Function\\ of Cosmic Structures: II Statistics }

\author[P. Monaco]
       {Pierluigi Monaco\\
        Scuola Internazionale Superiore di Studi Avanzati (SISSA), via  
        Beirut 4, 34014 -- Trieste, Italy \\
        Dipartimento di Astronomia, Universit\`a degli studi di Trieste, via 
        Tiepolo 11, 34131 -- Trieste, Italy\\
	Email: monaco@sissa.it}
 
\date{Received 1996 June 15}

\pagerange{\pageref{firstpage}--\pageref{lastpage}}
\pubyear{1996}

\begin{document}

\maketitle

\label{firstpage}

\begin{abstract}
The statistical tools needed to obtain a mass function from realistic
collapse time estimates are presented. Collapse dynamics has been
dealt with in paper I of this series by means of the powerful
Lagrangian perturbation theory and the simple ellipsoidal collapse
model. The basic quantity considered here is the inverse collapse time
\f; it is a non-linear functional of the initial potential, with a
non-Gaussian distribution. In the case of sharp $k$-space smoothing,
it is demonstrated that the fraction of collapsed mass can be
determined by extending to the \f\ process the diffusion formalism
introduced by Bond et al. (1991). The problem is then reduced to a
random walk with a moving absorbing barrier, and numerically solved;
an accurate analytical fit, valid for small and moderate resolutions,
is found. For Gaussian smoothing, the \f\ trajectories are strongly
correlated in resolution.  In this case, an approximation proposed by
Peacock \& Heavens (1990) can be used to determine the mass functions.
Gaussian smoothing is preferred, as it optimizes the performances of
dynamical predictions and stabilizes the \f\ trajectories.  The
relation between resolution and mass is treated at a heuristic level,
and the consequences of this approximation are discussed.  The
resulting mass functions, compared to the classical Press \& Schechter
(1974) one, are shifted toward large masses (confirming the findings
of Monaco 1995), and tend to give more intermediate-mass objects at
the expense of small-mass objects. However, the small-mass part of the
mass function, which depends on uncertain dynamics and is likely to be
affected by uncertainties in the resolution--mass relation, is not
considered a robust prediction of this theory.
\end{abstract}


\section{Introduction}

An important outcome of any cosmological model is the mass
distribution of those cosmic collapsed structures which are predicted
to form; this quantity is usually called mass function (hereafter MF)
or multiplicity function. The theoretical determination of this
quantity is difficult, as cosmological collapsed structures are the
sites of non-linear gravitational dynamics.  No exact analytical
solution of the non-linear collapse of a general self-gravitating
system is known. The first attempt to determine the number of
collapsed objects was made by Press \& Schechter (1974; hereafter PS);
to predict the collapse of a mass clump, they used a heuristic
argument based on the extrapolation of linear theory to the highly
non-linear regime, and on the spherical collapse model, whose
solutions are analytically known. Since PS, most works on the MF have
been based on similar dynamical arguments.

However, a number of dynamical approximations have recently been
developed. These approximations provide a reasonable description of
the collapse of a self-gravitating, pressure-less fluid up to caustic
formation, when the orbits of different mass elements cross each other
(orbit crossing, hereafter OC, or shell crossing). In a previous paper
(Monaco 1995, hereafter M95), the effects of non-spherical collapse
were analyzed in a PS-like approach; in that case, the fraction of
collapsed mass was identified with the probability of having suitable
initial conditions such as to make a mass element collapse. The
dynamical tools used in that case were the Zel'dovich approximation
(Zel'dovich 1970) and the homogeneous ellipsoid collapse model.

This paper is the second in a series in which a new theory for the MF
of cosmic structures is constructed. The idea, contained in M95, of
constructing an MF based on realistic collapse dynamics, is the basis
of the whole theory. The first paper of this series (Monaco 1996,
hereafter paper I) develops the dynamical tools needed to obtain an
MF. As already noted by M95, the MF dynamical problem (in its fluid
limit) is intrinsically Lagrangian, in the sense that it is best faced
within a Lagrangian fluidodynamical framework. Thus, all the tools of
the Lagrangian formulation of gravitational dynamics can be used: the
Zel'dovich approximation, Lagrangian perturbation theory (see, e.g,
Bouchet et al. 1995; Buchert 1994; Catelan 1995; complete references
are given in paper I), and the ellipsoidal collapse model. In paper I,
all these dynamical approximations are analyzed, with the following
results:

\begin{enumerate}
\item As in M95, the collapse of a mass element is identified
with the OC instant; this definition has been amply discussed.
\item Lagrangian perturbations are applied to smoothed versions
of the initial potential; it is assumed that smaller scales do
not influence the collapse significantly.
\item The Lagrangian series, up to third order, converges in
predicting the collapse of a homogeneous ellipsoid.
\item As a consequence, the collapse time of a homogeneous
ellipsoid can be estimated in an easy and fast way by means of the
third-order Lagrangian series, with a correction for quasi-spherical
collapses.
\item Ellipsoidal collapse can be seen as a particular
truncation of the Lagrangian series, when all the more than second
derivatives of the initial peculiar gravitational potential are
neglected.
\item In the general case of a Gaussian field with scale-free power
spectrum, the Lagrangian series is shown to converge in predicting the
collapse of a mass element, when fast-collapsing mass elements,
representing at least 10 per cent of the mass, are taken into
account. Convergence is valid at a qualitative level for about 50 per
cent of the mass.
\item The homogeneous ellipsoidal collapse correctly
reproduces the collapse time prediction of the Lagrangian series in
the same convergence range.
\end{enumerate}

The main quantitative outcome of paper I is the probability
distribution function (hereafter PDF) of the inverse collapse times
\f\ (in the spherical collapse case \f\ is just proportional to the
density contrast). These are preferred to the collapse times as their
distribution is better behaved: the inverse collapse times are large,
but of order one, for fast collapsing points, and become smaller and
smaller for slowly collapsing ones; non-collapsing points have
vanishing or negative \f\ values. 

This paper, paper II, faces the problem of finding an MF from the PDF
of the inverse collapse times. The statistics needed to `count' the
objects, which form according to the dynamical predictions used in
paper I, is developed, and the resulting MFs are analyzed and
commented.  A comparison of the whole theory to N-body simulations is
underway (paper III).  This paper is organized as follows: Section 2
contains an overview of the statistical procedures which have been
developed by previous authors to obtain an MF; this is useful so as
better to define the strategy for the statistical procedure to
develop.  In Section 3 the fraction of collapsed mass, as a function
of the resolution, is determined for sharp $k$-space filtering. This
quantity can be determined by solving a multi- (infinite-) dimensional
diffusion process with an absorbing barrier; however, it is
numerically shown that the same solution can be found by finding the
first upcrossing rate of the \f\ process alone, as if \f\ were a
Markov process; the precise behaviour of the \f\ process as a function
of resolution is left as an open problem.  Then, the determination of
the MF can be reduced to a diffusion problem with a moving absorbing
barrier. Numerical solutions and accurate analytical approximations
for the fraction of collapsed mass are presented. In Section 4 the
Gaussian smoothing case is analyzed. In this case the collapse time
trajectories present strong correlations in resolution, which
considerably complicate the problem. The results obtained from
numerical simulations of individual trajectories are successfully
compared to a reasonable approximation proposed by Peacock \& Heavens
(1990).  The passage from the resolution variable to the mass variable
is discussed in Section 5: a `missing piece' is identified, related to
the mass distribution of the extended collapsed structures. Section 6
contains a summary of the main results of the paper and some final
comments.

\section{Statistics and the MF: an Overview}

Once a collapse condition is given for every point, and its
statistical properties, namely its PDF, are known, a statistical
framework has to be developed, to `count' the number of collapsed
structures. The main statistical procedures proposed in the literature
are reviewed in the following.  Throughout the paper the following
notation will be used: the variable $\mres=\sigma_\delta^2$, equal to
the mass variance, will be used as the resolution variable, \f\ will
denote the inverse collapse time, and its PDF will be denoted as \pf;
in this Section, \f\ is generally equal to $\delta_l/\delta_c$, where
$\delta_l$ is the linearly extrapolated density contrast and $\delta_c$
is a threshold. The growing mode $b(t)$ will be used as the physical
time variable; it is demonstrated in paper I that this choice makes it
possible to parameterize out the dependence of dynamics on the
background cosmology with a great accuracy. The integral MF, i.e.  the
fraction of collapsed mass, will be denoted by \ot\ or \om, whether it
is considered as a function of the resolution \res\ or of the mass
$M$.  Finally, the differential MF, the number of objects per unit
volume and unit mass interval, will be denoted by \mf, while \mft\
will denote the \res\ derivative of \ot\ (see Section 5 for the
relation between the two $n$ functions).

In the seminal PS paper, a point was supposed to collapse whenever its
linearly extrapolated density contrast, smoothed at a scale
$R(\Lambda)$ with some filter, exceeded a given threshold $\delta_c$
(equal to 1.69 if the spherical collapse model was invoked). Thus, the
fraction of collapsed mass at a given resolution was:

\be \mot=\int_{F_c}^{\infty} \mpf dF \; ; \label{eq:ps} \ee

\noindent (\fc=1 if $F=\delta_l/\delta_c$). To get the MF as a function
of the mass, $M$ was set equal to the mass contained by the smoothing
filter; in the top-hat filter case, $M=4\pi\overline {\rho}R
(\Lambda)^3/3$. When the filter is not a top-hat in real space, one
can again obtain a relation between \res\ and $M$, even though its
physical meaning is less clear (see, e.g., Bond et al. 1991).

One of the weaknesses of this statistical approach was immediately
clear: for any threshold $\delta_c$ of order one, the integral in
equation (\ref{eq:ps}) tends to 1/2 as \res\ becomes very large, so
that just one half of all the mass in the Universe is predicted to
collapse; PS overcame this normalization problem by multiplying their
MF by a `fudge' factor of 2. This 1/2 factor is a signature of linear
theory, which predicts that only initially overdense regions (one half
of the total mass) are able to collapse. On the other hand, the lack
of normalization of the PS MF is caused by the incorrect way of
counting collapsed points.  As a matter of fact, a collapse prediction
is given to each point {\it for any resolution} \res; in other words,
a whole trajectory in the \f--\res\ plane is assigned to each point.
At small \res\ values (large $R$), all trajectories lie below the
threshold, but when \res\ grows the trajectories can upcross or
downcross the threshold.  The first upcrossing is related to the
collapse of the point at that scale; any further downcrossing is
meaningless: a point which is collapsed at a scale $R$ (and is
included in a region of size $\geq R$) cannot be considered as not
collapsed at a smaller scale $R'<R$ (and not included in any collapsed
region of size $\geq R'$). This fact was first recognized and solved
by Epstein (1983), then by Peacock \& Heavens (1990; hereafter PH) and
by Bond et al. (1991; hereafter BCEK). To correct for this {\it
cloud-in-cloud} problem, the integral MF has to be related to the
probability \pfu\ that a trajectory has never upcrossed the threshold:

\be \mot=1-\int_{-\infty}^{F_c} \mpfu dF \; .\label{eq:psu} \ee

The determination of \pfu\ is generally difficult: \pfu\ is the
conditional probability of \f\ being $>F_c$ at \res, given $F<F_c$ at
all smaller \res\ values, so that all the N-point correlations of \f\
at different \res\ have to be considered. The problem is greatly
simplified if the filter function is a top-hat in the Fourier space
(sharp $k$-space filtering, hereafter SKS; see BCEK).  In this case,
provided the initial density field is a Gaussian process, independent
modes are added as the resolution changes, and the trajectories in the
\f--\res\ plane are random walks (\f\ is a Markov process, in
particular a Wiener process; see Section 3 for further details).
Thus, the first upcrossing problem is equivalent to a random walk with
a (fixed) absorbing barrier.  The PDF of the trajectories obeys a
Fokker-Planck equation (see equation \ref{eq:genfp}):

\be \frac{\partial}{\partial \mres} \mpfu = \frac{1}{2} \frac{\partial^2}
{\partial F^2} \mpfu \; ,\label{eq:fpa} \ee

\noindent with the boundary condition $P_F^{\rm noup}(F_c,\mres)=0$. 
The solution is:

\bea \lefteqn{\mpfu dF =} \label{eq:fpsol} \\
&& \frac{1}{\sqrt{2\pi\mres}} \left[ \exp \left( -\frac{F^2}{2\mres}
\right) - \exp \left( -\frac{(F-2F_c)^2}{2\mres} \right) \right]dF \; .
\nonumber \eea

\noindent With this solution, the resulting MF is the original PS one,
including the `fudge' factor 2:

\be n(\mres) d\mres = F_c \exp(-F_c^2/2\mres)/\sqrt{2\pi\mres^3}d\mres\; .
\label{eq:psmf} \ee

\noindent If the filter is not SKS, the \f\ trajectories are not
random walks (\f\ is no longer a Markov process); they are affected by
strong correlations, and no Fokker-Planck equation can be written for
their PDF. In this case, the differential MF reduces to that of PS
{\it without} the factor of 2 at large masses, and has a different
slope at small masses; this was shown both by PH and by BCEK. Non-SKS
filters are very difficult to deal with in this framework, as all the
N-point correlation functions are relevant in determining the
statistical properties of the trajectories. However, PH found a
reasonable and successful way to approximate the \pfu\ distribution;
it will be described in Section 4.

The powerful and elegant diffusion formalism has been used to find the
merging histories of dark-matter clumps, as the solutions of a
two-barrier diffusion problem (BCEK; Lacey \& Cole 1993); Bower (1991)
obtained the same results as extensions of the original PS work,
without using the diffusion formalism.  Lacey \& Cole (1994) made a
number of empirical choices to optimize the adherence of their
formulae to N-body simulations: they took the SKS solution, which
coincides with the original PS solution, and is easier to deal with,
and let two parameters vary, namely the threshold parameter $\delta_c$
and the mass assigned to the filter; with these choices, they
succeeded in fitting the abundance and the merging histories of
simulated friends-of-friends dark halos.

Another important problem of the PS statistical procedure was outlined
by Blanchard, Valls-Gabaud \& Mamon (1992) and Yano, Nagashima \&
Gouda (1996) (even though the same problem had already been faced in
Epstein 1983). If the spherical top-hat model is used to predict the
collapse time, and if top-hat smoothing is consistently used, then the
density contrast in a point has to be interpreted as a mean over a
spherical volume of radius $R$. As a consequence, a collapse
prediction has to be considered as relative not simply to the point,
but to the whole spherical region which surrounds it. The collapse
condition has then to be changed to the following: a point collapses
if it is embedded in a collapsing region, even though it is not at its
center (and then its smoothed density contrast can be smaller than the
threshold). In paper I this kind of reasoning has been called {\it
global}  interpretation of the collapse time.  This new collapse
condition considerably complicates the statistical problem; Blanchard
et al. (1992) have shown this to cause a flattening of the MF with
respect to the PS one, while Yano et al. (1996) have faced the problem
by explicitly introducing the two-point correlation function in the
collapse condition.

The PS procedure was extended in a number of other papers. In
particular, Lucchin \& Matarrese (1988) extended the PS formalism to
non-Gaussian density fields and Lilje (1992) to general cosmologies.
Porciani et al. (1996) tried to introduce non-Gaussianity in the
diffusion framework by reflecting all the trajectories crossing
$\delta_l=-1$, to avoid unphysical negative densities; in this way they
found an intriguing cutoff of the MF at small masses.  The same
dynamical content as in the PS approach is present in the block model
of Cole \& Kaiser (1989) and in the Monte Carlo approach of Rodriguez
\& Thomas (1996).

In the PS framework, the relevant objects to be considered are the
sets of points whose \f\ upcrosses a given threshold; these sets are
usually called {\it excursion sets}. Alternatively, one can suppose
that the forming structures are not connected to general excursion
sets, but to the peaks of the initial density contrast (see, e.g.,
Bardeen et al. 1986). Then, counting structures is reduced to counting
the number of peaks above a given threshold. With respect to the
excursion set approach, the peak approach has a great advantage,
namely that the extended structures which are to be counted are well
defined, clearly connected to the peaks of the density field. However,
the peak approach has a number of disadvantages: (i) going from
excursion sets to peaks greatly complicates the formalism, as the peak
constraint is much stronger than a simple threshold constraint; as a
consequence, an analytical determination of the number of peaks is
hopeless if \f\ is not a Gaussian or closely related process (except
for high peaks; see Catelan, Lucchin \& Matarrese 1988); (ii)
obtaining an MF from the number of peaks requires an estimate of the
mass associated to a peak. Different reasonable choices lead to
different MFs (e.g., Ryden 1988; Colafrancesco, Lucchin \& Matarrese
1989; PH; Cavaliere, Colafrancesco \& Scaramella 1991); (iii) it is
very difficult to solve the `peak-in-peak' problem; this has been done
by PH in a heuristic way, then by Appel \& Jones (1990) and Manrique
\& Salvador-Sole (1995, 1996), and by Bond \& Myers (1996) within their
peak-patch theory; (iv) the peaks of the initial density field are
generally not the first points to collapse; this has been shown by
means of both theoretical arguments (see, e.g., Shandarin \&
Zel'dovich 1989: in the Zel'dovich approximations, structures are
connected with the peaks of $\lambda_1$, the largest eigenvalue of the
deformation tensor) and numerical simulations (Katz, Quinn \& Gelb
1993; van de Weigaert \& Babul 1994).

A different approach, pioneered by Silk \& White (1978) and Lucchin
(1989), was used by Cavaliere and coworkers; see Cavaliere, Menci \&
Tozzi (1994) for a recent review and paper I for further comments. In
their case, the existence of an MF is implicit, and kinetic evolution
equations are given for its evolution; this is at variance with the PS
and related approaches, where the MF is obtained from the evolution of
the density contrast field.  Besides, in their application of the
Cayley tree formalism to the adhesion approximation (Cavaliere, Menci
\& Tozzi 1996; see also Vergassola et al. 1995), the objects were
identified with shocks in the collapsing medium.

To determine the MF from the statistics of the \f\ process, the
excursion set approach has been chosen. It will be shown in the
following that, in the SKS smoothing case, despite the complicate
non-Gaussian nature of \f, the MF problem can be recast in terms of a
random walk with a moving absorbing barrier, while in the Gaussian
filter case the useful approximation proposed by PH applies. A peak
approach applied to the process \f\ would better take into account the
complex geometry of the collapsed regions in the Lagrangian space,
but, as mentioned before, it is intractable from the analytical point
of view.

Some remarks about the \f\ process, already outlined in paper I, are
worth stressing again:

\begin{enumerate}
\item As \f\ is (the inverse of) a time, any threshold \fc\ in this
theory simply specifies the time at which the MF is examined; there is
no free $\delta_c$ parameter.

\item Smoothing is necessary because of the truncated nature of the
dynamical approximations used. Thus the shape of the filter has to be
chosen in order to optimize the performances of the dynamical
predictions; usually Gaussian filters are preferred.

\item The dynamical predictions are strictly {\it punctual}; in other
words, a point collapses only if it is predicted to collapse (at a
given scale), not if a neighboring collapsing point is able to involve
it in its collapse.
\end{enumerate}

Point (i) is a result of the more detailed dynamics contained in this
MF theory. Point (ii) implies that, even though we still have some
freedom in choosing the shape of the filter, the Gaussian filter case
has to be regarded as the most ``physical'' one. Point (iii) is
connected to the discussion about the global interpretation of
collapse times: in the Lagrangian perturbation case dynamical
predictions are punctual, then the complications introduced by the
global interpretation are avoided. Moreover, spherical symmetry of
collapsed regions in Lagrangian space is not imposed, as it happened
with the global interpretation of spherical collapse time.  On the
other hand, the exact size of the collapsed regions cannot be so
easily determined as in the spherical case: as a consequence of the
coherence of the underlying initial field, a point collapsed at a
scale $R(\Lambda)$ will be part of a structure of typical size $R$ (or
larger), but the exact determination of its size requires knowledge of
the space correlations of the process \f. In other words, the
difficulties skipped in the diffusion problem, thanks to the punctual
interpretation of the collapse time, are found again in the $\Lambda
\rightarrow M$ transformation. This will be analyzed in Section 5.

Before going on, it is worth commenting on the physical meaning of the
absorbing barrier in the Lagrangian dynamical context. The nature of
the dynamical prediction is such that most mass is predicted to
collapse at small scales (92 per cent according, for instance, to the
Zel'dovich approximation); the exact number is not easy to determine,
as the behavior of the strongest underdensities is not well predicted
by Lagrangian perturbation theory (see paper I). Anyway, it is
unlikely that about 10 per cent of mass in the strongest
underdensities is going to affect the MF in any observable mass
range. Thus, the diffusion formalism is not needed here just to ensure
a correct normalization, which is more or less guaranteed, but to
solve the cloud-in-cloud inconsistency of the original PS procedure,
which in this context has the following meaning.  As the power
spectrum has no (or a very small) intrinsic truncation, a collapse
prediction is assigned for every resolution to every point; these
collapse predictions all have, in principle, the same
validity. Recalling that collapse means to be enveloped in an OC
region, it is natural to expect any point to be in OC at a small
enough scale, and to be in single-stream regime at a large enough
scale. The PS statistical procedure would suffice to find \mft\ if the
transition to OC occurred only once, i.e. if the process \f\ never
downcrossed the \fc\ barrier.  A downcrossing of \f\ has the following
meaning: a point is in OC at a large scale, but it is not at a smaller
scale; this appears contradictory, as a collapsed structure does not
contain non-collapsed subclumps. To overcome this inconsistency, it is
assumed that OC at one scale implies OC at all smaller scales. This
corresponds to absorbing the trajectories of the \f\ process that
upcross the \fc\ barrier.

\section{Sharp $k$-space Smoothing}

The quantity $F(\Lambda)$ is a non-linear and non-local functional of
the initial (peculiar rescaled) potential $\varphi(\mq;\Lambda)$.
Then, it is not trivial to infer the properties of the \f\ process as
a function of \res. In the following, it will be shown that the multi-
(infinite-) dimensional process, defined by all the \f\ and \pot\
values in all the points, is a Markov process, so that the statistics
of first upcrossing of \f\ can in principle be found by solving a
multi- (infinite-) dimensional Fokker-Planck (hereafter FP) equation.
It will also be shown that the same first upcrossing statistics can be
reproduced by studying a (one dimensional) Markov process with the
same PDF as the \f\ process. 

It is opportune, before going on, to introduce a number of definitions
which will be necessary in the following; a more detailed presentation
can be found, for instance, in the textbooks by Arnold (1973) and
Risken (1989).

Let $\xi(\mres)$ be a real stochastic process. Let's denote by $x$ the
value that the process $\xi$ takes at the resolution \res. The process
$\xi$ is said to possess the {\it Markov property} if its history at
resolutions $>\mres$ is determined by its value $x$ at \res,
independently of its past history; in other words, a process possesses
the Markov property if past and future are independent once the
present is known.

A Markov process is called a {\it diffusion process}\footnote{Note
that some authors reserve the name diffusion only to those processes
which have constant drift and diffusion coefficients.} if its
trajectories are continuous and if the quantities $D^{(1)}$ and
$D^{(2)}$ exist:

\bea \langle \xi(\mres+d\mres)-\xi(\mres) \rangle &=& D^{(1)} d\mres 
+ {\cal O} (d\mres^2) \label{eq:moments} \\
\langle (\xi(\mres+d\mres)-\xi(\mres))^2 \rangle &=& D^{(2)} d\mres 
+ {\cal O} (d\mres^2)\; . \nonumber  \eea

\noindent
The coefficients $D^{(1)}$ and $D^{(2)}$ are called {\it drift} and
{\it diffusion} coefficients.  The PDF $P_x(x,\mres)$ of a diffusion
process obeys a FP equation:

\be \frac{\partial}{\partial \mres}P_x(x,\mres) = \left[ -
\frac{\partial}{\partial x} D^{(1)}(x,\mres) + \frac{1}{2}
\frac{\partial^2}{\partial x^2} D^{(2)}(x,\mres) \right] P_x(x,\mres) \; . 
\label{eq:genfp} \ee

A diffusion process with $D^{(1)}=0$ and $D^{(2)}=1$ is called a {\it
Wiener process}, ${\rm W}(\mres)$. Its increments are uncorrelated:
$\langle d{\rm W}(\mres)d{\rm W}(\mres')\rangle=\delta_D(\mres-
\mres')$ ($\delta_D$ is the Dirac delta function), while $\langle {\rm
W}(\mres){\rm W}(\mres')\rangle= \min(\mres, \mres')$.  The transition
probability from $\mres$ to $\mres'$ is a Gaussian with variance
$(\mres-\mres')$, and the PDF at resolution $\mres$ is a Gaussian with
variance $\mres$ and zero mean.  All these definitions can be easily
extended to the case of multi-dimensional processes.

In order to simplify the notation, in the following a stochastic
process and its numerical values, although different mathematical
objects, will be denoted by the same symbol.

\subsection{A FP Equation for \pf}

It is useful to analyze first, as an example, the simple case of
Zeldovich approximation in 2D. Initial conditions are given by the
three independent component of the first-order deformation tensor (a
symmetric 2D matrix), $w_1$, $w_2$ and $w_3$, the third one being the
non-diagonal component.  The two eigenvalues of the deformation
tensor, $\lambda_1$ and $\lambda_2$, are simply:

\be \lambda_{1,2} = \frac{1}{2}\left(w_1+w_2\pm\sqrt{(w_1-w_2)^2+4w_3^2}
\right)\; , \label{eq:eigdd} \ee

\noindent where the $+$ sign corresponds to the largest eigenvalue
$\lambda_1\equiv F$. 

The three $w_i$ components, being linearly connected to the initial
potential \pot, are correlated Gaussian variables and, as functions of
\res, Wiener processes, with nonvanishing mutual correlations at fixed
resolutions. Such processes constitute a 3D diffusion process; the
evolution of their joint PDF is then controlled by a 3D FP equation
with vanishing drift and constant diffusion ($D^{(2w)}_{ij}$)
coefficients.  Collapse is assumed to have place if:

\be F\equiv \lambda_1(w_1(\mres),w_2(\mres),w_3(\mres)) > F_c\; . 
\label{eq:barrier} \ee

\noindent 
Then, the first upcrossing rate problem can be formulated in term of a
3D diffusion with a complicate absorbing barrier.

On the other hand, the system $\{\lambda_1,\lambda_2,w_3\}$ can be
obtained by means of a non-linear transformation of the $\{w_i\}$
system; it is easy to verify that such transformation is invertible
and twice differentiable in any point, except in the line $w_1=w_2$
and $w_3=0$, a set of zero measure which is explicitly avoided by
trajectories (as in the 3D case; see the joint PDF of eigenvalues,
e.g., in M95).  In this case, the $\{\lambda_i\}$ (with $\lambda_3
\equiv w_3$) system is a Markov process, and, more precisely, a
diffusion process (Arnold 1973; Risken 1989). The drift and diffusion
coefficients can be found by means of the following transformation
rules:

\bea D^{(1)}_i = \frac{\partial^2 \lambda_i}{\partial w_j \partial w_k} 
D^{(2w)}_{jk} \\
D^{(2)}_{ij} = \frac{\partial \lambda_i}{\partial w_k} 
\frac{\partial \lambda_j}{\partial w_l} D^{(2w)}_{kl} \; . 
\nonumber \label{eq:transf} \eea

\noindent
Then, the problem can be reformulated by means of a more complicate FP
equation with the simple barrier $F=F_c$.

The same considerations can be generalized to any sufficiently
well-behaved inverse collapse time \f\ in 3D.  Let $\varphi(\mq)$ be a
Gaussian process in the Lagrangian space \q\ (see paper I for the
definition of Eulerian and Lagrangian coordinates).  \pot\ is the
initial peculiar rescaled gravitational potential, defined by the
equation:

\be \nabla^2 \varphi(\mq) = \delta_l (\mq)\; , \label{eq:poi} \ee

\noindent
where $\delta_l(\mq)=\delta_0(\mq)/b_0$ is the linear density contrast
($\delta_0$ is the initial density contrast and $b_0\simeq a_0$ is the
initial growing mode, nearly equal to the initial scale factor). This
Gaussian process is assumed to have power at all relevant scales, up
to a very small cutoff.  From the field \pot\ a hierarchy of smoothed
fields is obtained:

\be \varphi(\mq) \rightarrow \varphi(\mq;\mres) = \varphi(\mq) *
W(\mq,\mres) \; ,\label{eq:smooth} \ee

\noindent 
where $*$ denotes a convolution, and $W(\mq,\mres)$ is the SKS filter,
which in the Fourier space is $\theta(1-kR(\mres))$.

The inverse collapse time $F(\mq,\mres)$ of a point is a functional
acting on the \pot\ potential:

\be F(\mq,\mres) = b_c^{-1}(\mq,\mres)= {\cal F}_\mq[\varphi(\mq',\mres)]\; . 
\label{eq:func} \ee

\noindent 
It is important to note that the functional ${\cal F}$ does not act
directly on the \res\ variable; in other words, it does not shuffle
information coming from different resolutions. In many cases it is not
possible to obtain the explicit form of this functional (see paper I
for full details).  Anyway, it will always be assumed that it is twice
differentiable with respect to the \pot\ values, except in some
possible degenerate cases, which constitute a set of zero measure in
Lagrangian space.

It is convenient, in order to handle the functional ${\cal F}$
introduced above, to consider discrete spaces; all the considerations
which follow will be valid in the continuum limit.  Let's then
consider the (Lagrangian) space divided into a large number $N$ of
points $\{\mq_i\}$:

\bea 
F(\mq,\mres) & \rightarrow & F(\mq_i,\mres) = F_i(\mres) \nonumber\\
\varphi(\mq,\mres) & \rightarrow &\varphi(\mq_i,\mres) = \varphi_i(\mres) 
\label{eq:disc} \\
{\cal F}_\mq[\varphi(\mq',\mres)] & \rightarrow & {\cal F}_{\mq_i}
(\{\varphi(\mq_j,\mres)\}) = {\cal F}_i(\{\varphi_j(\mres)\})\; . \nonumber
\eea

\noindent 
In this way the functional becomes an ordinary (non-linear) function
of the $N$ resolution-dependent variables $\{\varphi_i(\mres)\}$.
 
The evolution equations of the $\{\varphi_i\}$ variables are those of
a vector Wiener process. They can be written as follows:

\be d\varphi_i = f_{ij}(\mres)d{\rm W}_j \label{eq:wiener} \ee

\noindent (summation over repeated indexes is meant), where $\{d{\rm
W}_j\}$ are $N$ independent Wiener processes. The $f_{ij}(\mres)$
coefficients are such as to give the correct variances and spatial
correlations of \pot.  Note the use of the differential notation in
equation (\ref{eq:wiener}), which is common in stochastic mathematics,
as time derivatives are ill-defined in this context.

It is possible to find evolution equations for the functions
$F_i(\mres)$. With a chain-rule differentiation, assuming twice
differentiability, the following system is obtained:

\bea dF_i &=& \frac{\partial{\cal F}_i}{\partial \varphi_j} (\{\varphi\})
d\varphi_j \label{eq:system} \\ d\varphi_i &=& f_{ij}(\mres) d{\rm W}_j\; . 
\nonumber \eea

\noindent
This is a non-linear Langevin system. As mentioned above, this implies
that the whole system $\{F,\varphi\}$ is a Markov process; provided
the regularity conditions given above (eq. \ref{eq:moments}) are
satisfied, the system is a diffusion process (see Arnold 1973 and
Risken 1989 for a complete demonstration). Then, the evolution of
$P_{\{F,\varphi\}}$, the joint PDF of the \f\ and \pot\ values {\it in
all the points}, is given by a FP equation:

\be \frac{\partial P_{\{y_i\}}}{\partial \mres} = \left[ - \frac{\partial}
{\partial y_i} D^{(1)}_i(\{y_i\}) + \frac{1}{2} \frac{\partial^2}{\partial 
y_i \partial y_j} D^{(2)}_{ij}(\{y_i\}) \right] P_{\{y_i\}}\; , 
\label{eq:multifp} \ee

\noindent
where $y_i$ denote the components of the $\{F,\varphi\}$ vector. To
find the first upcrossing rate, e.g., of the $F_1$ process (evaluated
at $\mq_1$) above the barrier \fc, such equation has to be solved with
the constraint $P_{\{F,\varphi\}}(F_1=F_c,\ldots)=0$.

Provided the $\{\varphi\}\rightarrow\{F\}$ transformation is a well
defined and invertible one, the $\{F\}$ system is a Markov process by
itself, and the problem can be formulated in terms of the $\{F\}$
system only.

A solution of such a multi-dimensional problem, while revealing the
Markov nature of the problem, is not convenient in practice. A great
simplification would come from reducing such a multi-dimensional
problem to a one-dimensional one. As the 1-point PDF of the F process
alone is known, it is possible to construct a FP equation whose
solution is \pf; if \f\ were a Markov process, the constrained
solutions of the 1D FP equation would give the right solution for its
first-upcrossing rate.  However, there seems to be no simple reason to
conclude that \f\ is a Markov process: the components of a
multi-dimensional Markov process can possess not the Markov property
(see, e.g., the example given in Risken 1989, Section 3.5); moreover,
it is not possible to obtain a FP equation for $F=\lambda_1$ by
integrating out all the other variables in the FP equation
(\ref{eq:multifp}), as the drift and diffusion coefficients are not
constant and then cannot be taken out of the integrals.

The solution of the 1D problem can anyway be considered as an {\it
ans\"atz} to the true solution; it can be checked {\it a posteriori},
by means of direct numerical calculations, whether such {\it ans\"atz}
leads to a correct description of the exact first upcrossing
rate. This will be done in next Subsection, where it will be shown
that the 1D solution leads to an almost perfect description of the
first upcrossing rate problem. The numerical demonstration of this
fact is sufficient for the scope of this paper, but leaves the
question open of the reason of such behaviour, and of the possible
Markov nature of the \f\ process. Such complex problems will be
addressed in future work.

To construct a FP equation whose solution, with initial condition
$P_F(F,0)=\delta_D(F)$ and natural boundary conditions (\pf\ and its
\f-derivative vanish at infinity), is \pf (as given in paper I), it
is convenient to transform the \pf\ distribution to a Gaussian
distribution with variance \res, \px. The $x(F)$ transformation for
$\mres=1$ is given in paper I for the \f\ processes corresponding to
ellipsoidal collapse (hereafter ELL) and third-order Lagrangian
collapse (hereafter 3RD). To obtain the \pf\ distribution and the
$x(F)$ transformation at any \res, the following self-similarity
property can be used: $P(F/\sqrt{\mres}) dF/\sqrt{\mres} = {\rm
const}$; this is also valid for the \px\ distribution. Then,

\bea P_F(F,\mres)dF &=& \sqrt{\mres}\, P_F(F/\sqrt{\mres},1)dF
\label{eq:resc}  \\
x(F,\mres) &=& \sqrt{\mres}\, x(F/\sqrt{\mres},1)\; . \nonumber \eea

With this transformations, the unconstrained PDF of the transformed
quantity $x$ is virtually identical to the distribution of the linear density
contrast $\delta_l$ (this is true at the 1-point level; the N-point PDFs
of the transformed quantities $x$ at different points will generally
not be Gaussian multivariates). The FP equation which admits \px\ as a
solution is obviously that of a Wiener process:

\be \frac{\partial}{\partial \mres} \mpx = \frac{1}{2} \frac{\partial^2}
{\partial x^2} \mpx\; . \label{eq:fpx} \ee

\noindent 
Transforming back to the $F$ variable, the FP equation for the \pf\
distribution is obtained; its drift and diffusion coefficients are:

\bea D^{(1)}_F &=& - \left( \frac{\partial x}{\partial \mres} + 
\frac{1}{2} \frac{\partial^2 x}{\partial F^2} \left(
\frac{\partial x}{\partial F} \right)^{-2} \right) \left(
\frac{\partial x}{\partial F} \right)^{-1} \label{eq:ddcoefa}\\
D^{(2)}_F &=& \frac{1}{2} \left(\frac{\partial x}{\partial F} \right)^{-2}\; .
\nonumber \eea

\noindent If the transformation is linear, $x=aF+b\sqrt{\mres}$, then
the two coefficients become:

\bea D^{(1)}_F &=& -\frac{b}{2a\sqrt{\mres}} \nonumber\\
D^{(2)}_F &=& \frac{1}{2a^2} \; . \label{eq:ddcoefb} \eea

\noindent
However, it is convenient to work directly with the $x$ process, whose
FP equation is much simpler.

\begin{figure}
\centerline{
\psfig{figure=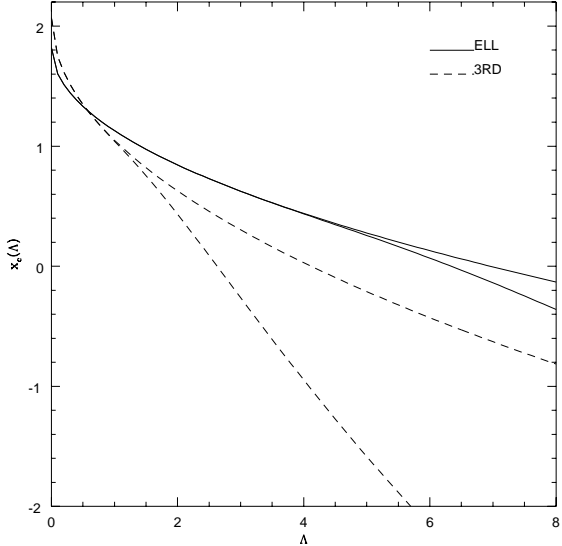,width=8cm}
}
\caption{The $x_c(\mres)$ barriers for the ELL and 3RD
predictions (heavy lines), together with their `linear' part (light
lines).}
\end{figure}

To find the distribution of the trajectories which have not upcrossed
a given threshold \fc, the FP equation for \f\ has to be solved with
the boundary condition $P_F(F_c,\mres)=0$. Transforming this condition
back to the $x$ variable, as the $F\rightarrow x$ transformation is
time-dependent, the absorbing barrier for the $x$ process will move as
\res\ grows. Then, the MF problem is reduced to a Wiener diffusion
problem with a moving absorbing barrier.

\subsection{The solution of the Moving Barrier Problem}

The diffusion problem with a fixed absorbing barrier, equivalent to
the solution of the FP equation (\ref{eq:fpx}) with boundary condition
$P_x(x_c,\mres)=0$, and with initial condition $P_x(x,0)=\delta_D(x)$,
has a solution that has long been known (Chandrasekhar 1943), which can
be obtained in the following way: the initial condition is changed to
$P_x(x,0)= \delta_D(x)-\delta_D(2x_c-x)$, i.e. a negative image is put in
a position symmetric with respect to the barrier. In the subsequent
evolution, the boundary condition is satisfied at any time by
symmetry. It is easy to see that the initial condition just shown
leads to equation (\ref{eq:fpsol}) as a solution (with $x$ instead of
\f). This solution formally turns negative beyond $x_c$; the true
solution is obviously null there. Note also that no meaningful
solution exists if $x_c<0$: all the trajectories are absorbed from the
start.

\begin{figure*}
\centerline{
\psfig{figure=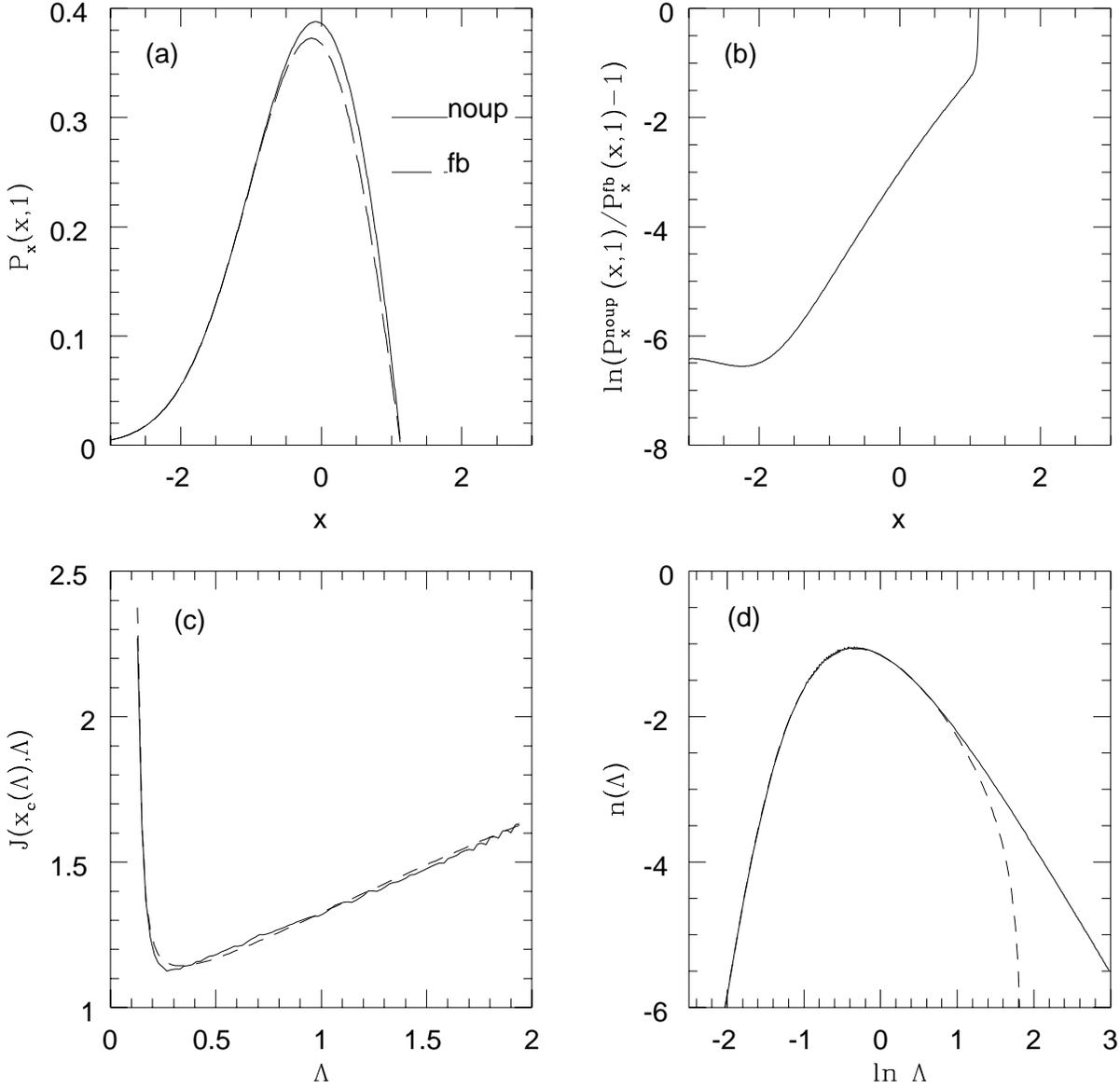,width=17cm}
}
\caption{ELL prediction, linear barrier, SKS
smoothing. a) \pxu\ and \pxb\ at \res=1. b) The logarithm of their
ratio minus 1. c) The $J$ correction factor, numerical result and
analytical fit. d) The resulting \mft\ with the analytical fit.}
\end{figure*}

\begin{figure*}
\centerline{
\psfig{figure=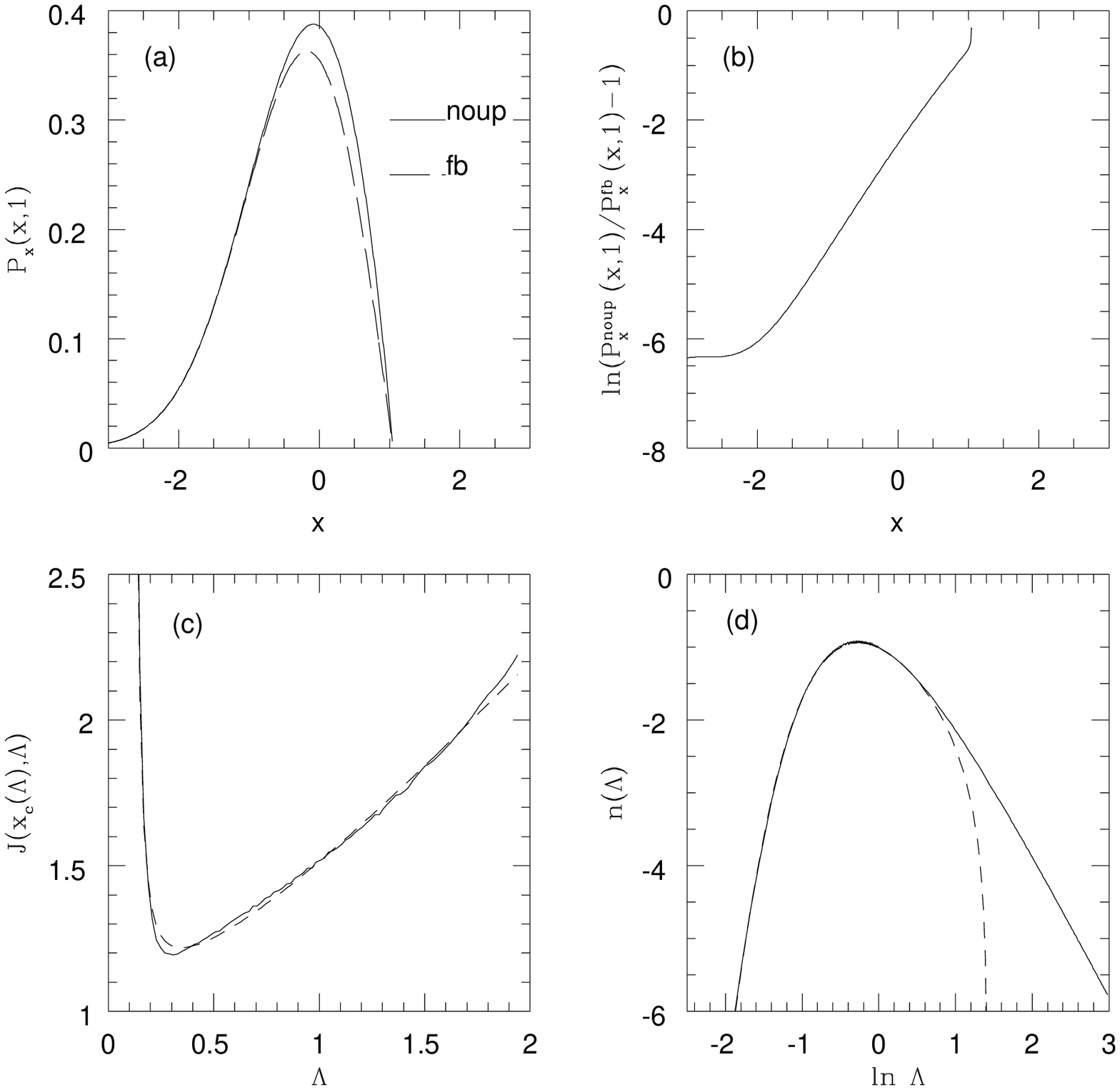,width=17cm}
}
\caption{The same as Fig. 2 with 3RD prediction. linear
barrier, SKS smoothing.}
\end{figure*}

In the moving barrier problem ($x_c=x_c(\mres)$) the image method
cannot be applied. In fact, a moving barrier problem is equivalent to
a diffusion problem with non-null drift and a fixed barrier; this is
the case for the $F(\mres)$ diffusion. In this case, any negative
image, put in a symmetric position with respect to the barrier, ought
to move in a specular way with respect to the positive component, in
order to ensure the PDF to be null at the barrier at any \res. Thus,
the image ought to move with a drift which is opposite in sign to the
drift of the positive component; as a consequence, its PDF would {\it
not} be a solution to the FP equation; it would be the solution to
another FP equation, with drift of opposite sign.

Appendix A contains a different form of the FP equation, in which the
barrier condition is implicitly contained in the drift and diffusion
coefficients. Moreover, a solution is given in terms of a path
integral.  Having found no simple analytical solution, a numerical
integration of the FP equation has been performed. Equation
(\ref{eq:fpx}) has been integrated by means of the Cranck-Nicholson
method, which consists in a finite-interval integration, stabilized by
an artificial numerical viscosity (see Press et al. 1992 for details).
The goodness of the result depends on the parameter $\alpha=\Delta
\mres/ 2(\Delta x)^2$; the result is stable for any $\alpha$, but the
small-scale features are better represented if $\alpha<1$. The
following finite intervals have been chosen: $\Delta x=7.5\;10^{-3}$,
$\Delta \mres =5.\; 10^{-5}$, which leads to $\alpha=0.444$; this is
quite adequate, as the resulting distributions are very smooth.

\begin{figure*}
\centerline{
\psfig{figure=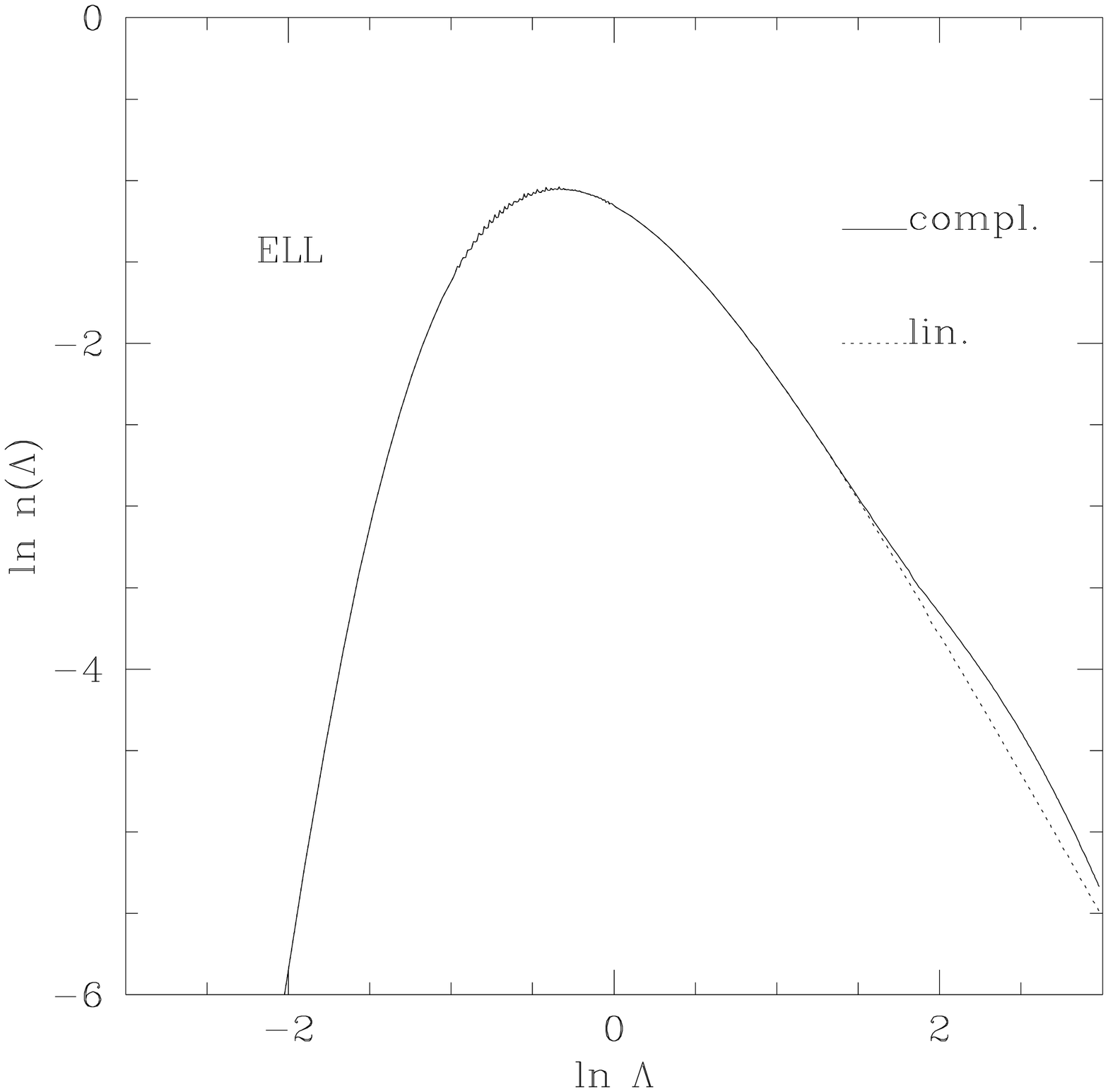,width=8cm}
\psfig{figure=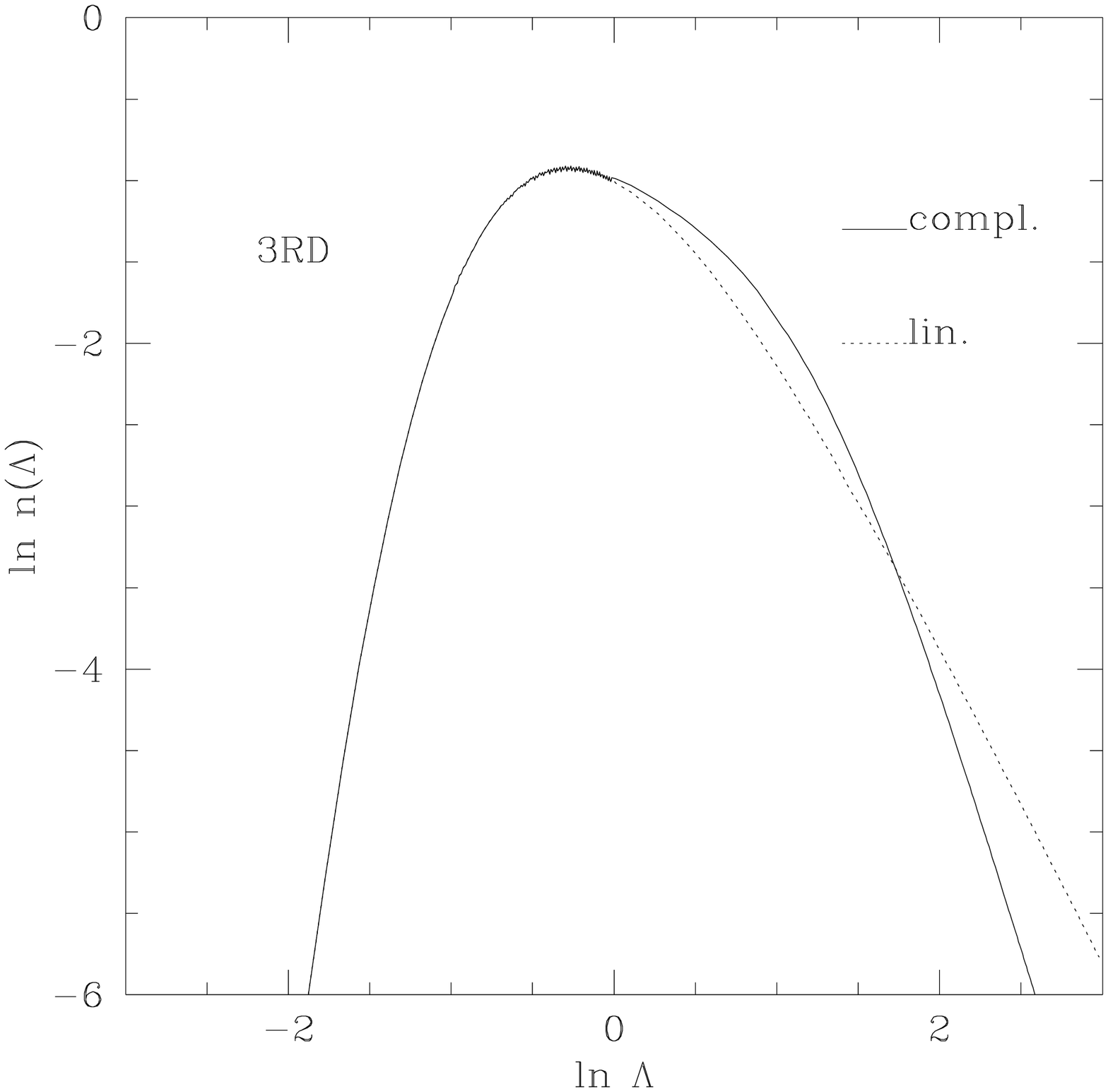,width=8cm}
}
\caption{The ELL and 3RD \mft\ curves: linear and complete
barriers.}
\end{figure*}

As shown in paper I, the $x(F)$ transformation is accurately linear in
\f\ when \f\ is larger than 1; this is true especially for ELL. In
this case, the absorbing barrier is:

\bea x_c(\mres) &=& 1.82 F_c - 0.69 \sqrt{\mres}\ \ {\rm (ELL)}
\label{eq:linxc}\\
     x_c(\mres) &=& 2.07 F_c - 1.02 \sqrt{\mres}\ \ {\rm (3RD)}\; . 
\nonumber\eea

\noindent Note that linear barriers are not linear in \res! In the
following, the $F_c=1$ barriers will be considered; any other $F_c$
value can be obtained by a \res\ rescaling: $\mres \rightarrow
F_c^2\mres$.  Fig. 1 shows the $x_c(\mres)$ barriers, based on both
linear and non-linear $F\rightarrow x$ transformations. The non-linear
ELL barrier is accurately reproduced by the linear one up to
moderately large \res\ values, while the non-linear 3RD barrier
significantly departs from the linear one beyond \res=1.

The numerical solutions found with linear barriers have been compared
with those found by simply inserting the $x_c(\mres)$ function into
the fixed barrier solution, equation (\ref{eq:fpsol}); these functions
are denoted by \pxb.  It turns out that the ratio between the two
solutions is accurately fit by:

\be \mpxu/\mpxb = {\cal J}(x,\mres) = 1+\exp(f(\mres)x+g(\mres))\; .
\label{eq:corr} \ee

\noindent Fig. 2a shows the numerical \pxu\ and the \pxb\ at \res=1,
for the linear ELL transformation; Fig. 2b shows the quantity
$\ln(\mpxu/\mpxb-1)$, which is accurately linear up to errors due to
numerical precision. Figs 3a and b show the same for the 3RD
prediction.

To determine the $f$ and $g$ functions, note that the resulting \pxu\
distribution is a solution of the FP equation (\ref{eq:fpx}); thus the
analytical fit has to reproduce its first derivative in t and the
second derivative in x. It is useful to obtain the \mft\ in terms of
the fixed-barrier solution \pxb\ and a correction term ${\cal
J}(x,\mres)$; then the $f$ and $g$ functions can be tuned so as to
give the correct ${\cal J}$.  The integral MF can be written as:

\bea \lefteqn{\mot = 1 - \int_{-\infty}^{x_c(\mres)}\mpxu dx =}
\label{eq:uno} \\
&& 1 - \int_{-\infty}^{x_c(\mres)} \mpxb {\cal J}(x,\mres) dx\; . \nonumber
\eea

\noindent \mft is then:

\bea 
n(\mres)=\frac{\partial\mot}{\partial\mres} 
&=& -\int_{-\infty}^{x_c(\mres)}\frac{\partial P_x^{\rm noup}}{\partial\mres} 
dx =\nonumber\\
&=& -\frac{1}{2}\int_{-\infty}^{x_c(\mres)}\frac{\partial^2 P_x^{\rm noup}}
{\partial x^2} dx = \label{eq:due}\\
&=& -\frac{1}{2} \frac{\partial P_x^{\rm fb}}{\partial x}{\cal J}(x_c
(\mres),\mres) = \nonumber\\
&=& \frac{x_c(\mres)}{\sqrt{2\pi \mres^3}}{\rm e}^{(-x_c(\mres)^2/2\mres)}
{\cal J}(x_c(\mres),\mres)\; .\nonumber \eea

\noindent The commutation between the integral and the time derivative
in the first passage is justified by the fact that the integrand
always vanishes at the upper integration limit. The second passage is
because of the FP equation (\ref{eq:fpx}). In the third passage, the
term with $x$-derivatives of ${\cal J}$ vanishes because \pxb\
vanishes at $x=x_c$.

This last expression has been compared to the numerical result in order to
tune the $f$ and $g$ functions. The best-fit expressions are:

\bea f(\mres) & = & -0.5  + 2.41\mres^{-1.08}\label{eq:ellfg}\\
     g(\mres) & = &  2.23 - 4.90\mres^{-1} \nonumber \eea

\noindent for the linear ELL transformation, and

\bea f(\mres) & = & -0.3  + 2.05\mres^{-1.13}\label{eq:trdfg}\\
     g(\mres) & = &  1.15 - 4.25\mres^{-1} \nonumber \eea

\noindent for the linear 3RD transformation.  Figs. 2c and 3c show the
numerical and proposed analytical corrections, equation
(\ref{eq:corr}), for the linear ELL and 3RD transformations. Figs. 2d
and 3d show the corresponding \mft; the agreement is excellent. Note
that this correction is valid as long as $x_c>0$; when the barrier
becomes negative (this fact has no relevant meaning), the
fixed-barrier solution vanishes, so that this procedure cannot be used
any more.

It may be useful to express the new MF in terms of the classical PS
one, with a free $\delta_c$ parameter. Writing the absorbing barrier
as $x_c(\mres)=\delta_c(1+(x_c(\mres)/\delta_c-1))$, the \mft\ can be
written as:

\bea 
n(\mres) &=& \frac{\delta_c}{\sqrt{2\pi\mres^3}}\exp\left( -\frac{\delta_c^2}
{2\mres}\right) \times \nonumber\\&& \left[ \frac{x_c}{\delta_c}\exp\left(
-\frac{2(x_c-\delta_c)+(x_c/\delta_c-1)^2}{2\mres}\right) {\cal J}\right]
\label{eq:corps}\\ &=& n^{PS}(\mres)\times{\cal I}(\mres)\; . \nonumber  \eea

\noindent The ${\cal I}$ correction factor has been defined in the
same way as in M95. 

The linear transformation is only an approximation, valid up to
moderately large \f\ values.  The complete transformation shows a
falling tail at low \f\ values, which corresponds to a pronounced peak
around \f=0.5 (Figs. 8a and b of paper I). The existence of this peak
is confirmed both by ELL and by 3RD, but the exact position is not
considered a robust feature; at those \f\ values the convergence of
the Lagrangian series is not guaranteed. If this falling tail is
modeled as in paper I, then the absorbing barrier becomes:

\bea x_c(\mres) &=& 1.82 F_c - 0.69 \sqrt{\mres} - \label{eq:comxc}\\&& 0.4 
\sqrt{\mres} ({\rm erf}(-7.5 F_c/\sqrt{\mres} + 1.75) + 1)\ \ {\rm (ELL)}
\nonumber\\ 
     x_c(\mres) &=& 2.07 F_c - 1.82 \sqrt{\mres} - \nonumber\\&& 0.75 
\sqrt{\mres} ({\rm erf}( -3  F_c/\sqrt{\mres} + 1.18) + 1)\ \ {\rm (3RD)}\; . 
\nonumber \eea

\noindent Fig. 4a shows the linear and complete ELL \mft, Fig. 4b
shows the same for 3RD. The small-mass\footnote{I freely use the word
mass in this context to indicate the large-mass (small \res) or
small-mass (large \res) part of the MF; the exact correspondence
between the two quantities is examined in Section 5.} part depends on
the details of the complete transformation, especially in the 3RD
case; in the ELL case the effect is modest even at rather large \res\
values.  As the details of the \pf\ distribution at low \f\ values are
not considered reliable, the low-mass part of the MF is not considered
a robust prediction of the theory.  This is especially true at very
large \res\ values: in this case the fit of the $x(F)$ transformation,
given in paper I and used to get equations (\ref{eq:comxc}), is not
accurate; on the other hand, that part of the \pf\ distribution is
very uncertain. Nonetheless, note that the non-linear element added in
the $x(F)$ transformation has the effect of enhancing the \mft\
function at moderate \res\ values, with a corresponding loss of
small-mass objects. This fact, which in some sense introduces a second
scale-length in the MF (the first, $M_*$, corresponding to the peak of
\mft), is somehow similar to the small-scale cutoff found by Porciani
et al. (1996) (their effect on the MF is much more dramatic). It is
thus confirmed that the introduction of dynamical non-Gaussianity can
lead to a large number of objects with intermediate masses, at the
expense of small-mass objects.

\begin{figure}
\centerline{
\psfig{figure=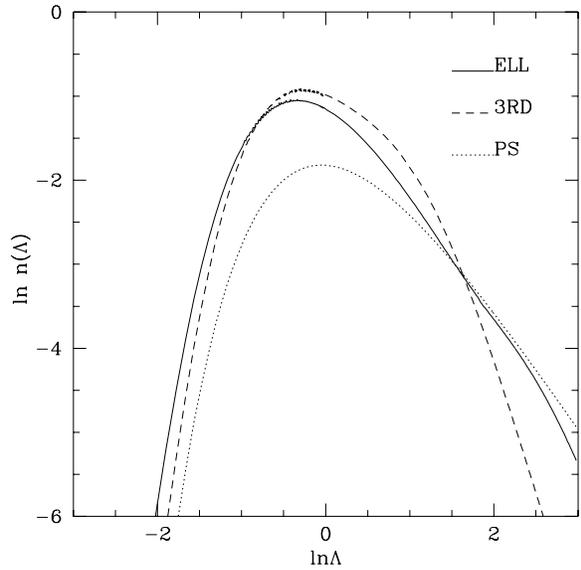,width=8cm}
}
\caption{SKS ELL and 3RD \mft\ curves, with complete
barriers. A PS curve is shown for comparison.}
\end{figure}

\begin{figure*}
\centerline{
\psfig{figure=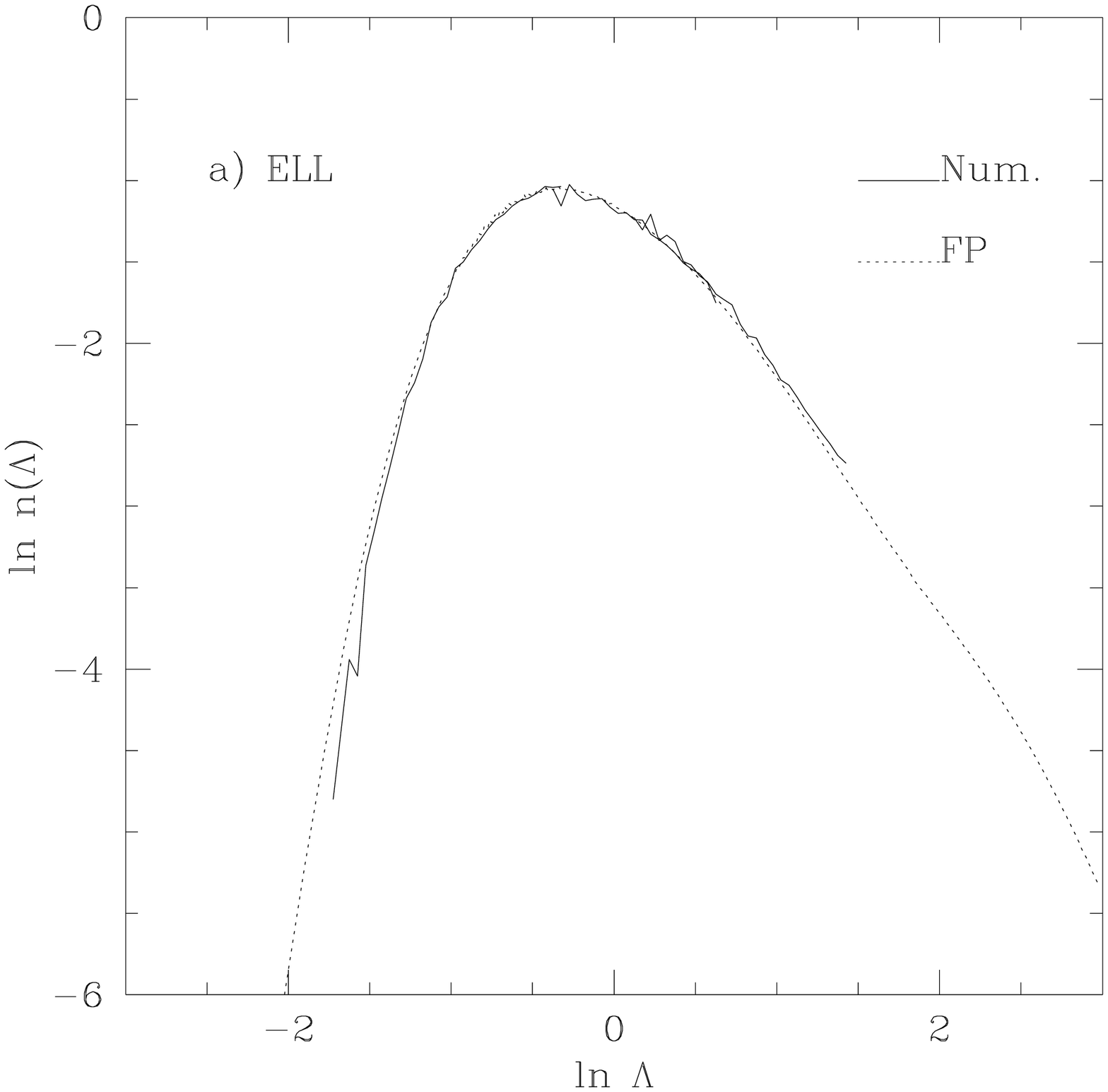,width=8cm}
\psfig{figure=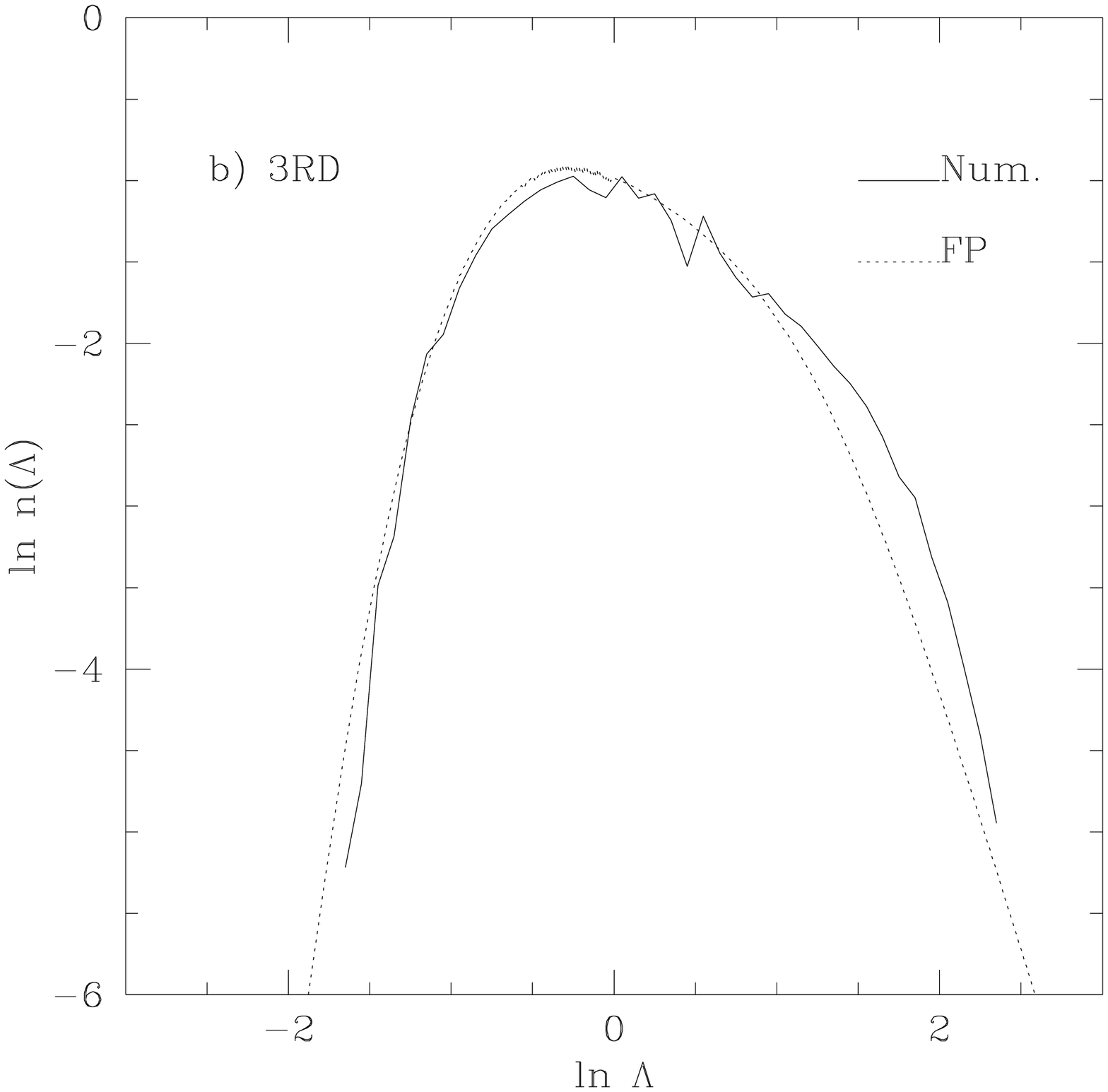,width=8cm}
}
\caption{SKS ELL and 3RD \mft\ curves, with complete barriers: FP and 
numerical solutions.} 
\end{figure*}

The (non-linear) complete ELL and 3RD \mft\ functions are shown in
Fig. 5, together with the original PS one (with the factor of 2).  ELL
and 3RD agree reasonably well at intermediate resolutions, while ELL
gives more large-mass objects, as 3RD slightly underestimates
quasi-spherical collapses (see paper I). On the other hand, their
small-mass behavior is dominated by the non-linear features of the
$x(F)$ transformation. Compared to the PS prediction, both ELL and 3RD
curves (i) show large-mass tails shifted toward large masses, (ii)
give about a factor ${\rm e}\simeq 2.7$ more objects than PS around
\res=0, and (iii) show steeper small-mass tails (especially
3RD). Point (i) is in agreement with the findings of M95; point (ii),
somehow worrisome (the PS curve is in agreement with N-body
simulations), will be solved by the use of Gaussian filters.

Finally, the solutions found by solving the moving barrier problem
have to be compared to the numerical solutions of the exact
multi-dimensional problem, as explained in section 3.1, to decide on
the validity of the FP solutions. The numerical solutions can be
obtained by means of the Monte Carlo calculations already presented in
paper I. Realizations of the initial potential are obtained in cubic
grids of 16$^3$ or 32$^3$ points; such realizations are smoothed by
means of a hierarchy of SKS filters, having care to sample the
$k$-space so as to follow exactly the spacing of the modes of the
cubic box. For any smoothing (116 for the 16$^3$ realizations, 464 for
32$^3$ ones), the collapse time is calculated for any point which has
not collapsed yet at smaller resolutions, and the upcrossing rate is
directly calculated. The resolution range of such calculations is
limited, so that, to cover a significant range, it is necessary to
perform different sets of realizations with different normalizations.
Three sets of 30 different 32$^3$ realizations, with total variances
0.8, 2 and 5, and scale-free power spectra with slope $-2$, have been
performed for the ELL case, while in the 3RD case, which is much more
time consuming, four sets of 60 16$^3$ realizations, with variances 
0.8, 2, 5 and 12 have been used.

Fig. 6 shows the comparison between the FP solutions and the direct
numerical calculations. In the ELL case, the agreement between the FP
and the numerical solutions is overall excellent. In the 3RD case,
very delicate from the numerical point of view, the agreement is good
at large masses, while the numerical curve is somehow below the FP one
by $\sim$10\% at intermediate masses. At small masses, $\ln\mres>1$, a
significant disagreement is visible; this is probably due to the fact
that the analytical expression proposed in paper I for the \pf\
distribution poorly fits the actual (numerical) distribution at small
\f\ values.  This gives an idea of the dependence of the MF at small
masses on the unreliable details of the dynamics of slowly collapsing
points; a more detailed fit of the \pf\ distribution is considered
unworthy. It is however noteworthy how the small-mass decrease of the
MF is emphasized by the numerical curve.

In conclusion, it has been demonstrated that the moving barrier
problem gives an accurate solution of the multi-dimensional problem
presented in Section 3.1. This fact permits on the one hand to solve
the MF problem in a reasonable way, just by extending the diffusion
formalism to the \f\ process, and on the other hand it sheds some
light on the nature of the \f\ process as a function of the
resolution; the question whether \f\ possesses the Markov property
will be faced by future work.

\section{Gaussian Smoothing}

The elegant diffusion formalism presented above is strictly limited to
SKS filtering. Any other filtering, which cuts the power spectrum in a
non-sharp way, creates strong correlations in the \f\ trajectories.
This can be seen in Fig. 7, where a sample of nine (Markov) SKS and
Gaussian trajectories are plotted (Gaussian trajectories are obtained
by smoothing the same SKS trajectories; a scale-free power spectrum
with $n=-2$ is used): while SKS trajectories are random walks, the
Gaussian trajectories are strongly correlated, with a significant
correlation length. This can be also seen as follows: for a Gaussian
process, the normalized correlation of the values of the process at
different resolutions behaves as follows:

\be \frac{\langle F(\mres)F(\mres+\Delta\mres)\rangle}{\sqrt{\langle 
F(\mres)^2 \rangle \langle F(\mres+\Delta\mres)^2\rangle}} \simeq
\mres \left( 1 -\frac{1}{2}\left( \frac{\Delta\mres}{\mres_c}\right)^2
\right)\; , \label{eq:dipend} \ee

\noindent i.e. it is constant to first order in $\Delta\mres$.  \lc\
is a spectrum-dependent coherence length, equal to $\mres
\sqrt{2(3+n)}$ for scale-free spectra; in general:

\be \mres_c = 2\mres\gamma(1-\gamma^2)^{-1/2}\; , \label{eq:lambdac} \ee

\noindent where $\gamma$ is a standard spectral measure (see Bardeen
et al. 1986); $\gamma=((n+3)/(n+5))^{1/2}$ for scale-free power
spectra. In the SKS case, the \lc\ scale vanishes, and the normalized
correlation linearly decreases for small \res\ variations.

\begin{figure}
\centerline{
\psfig{figure=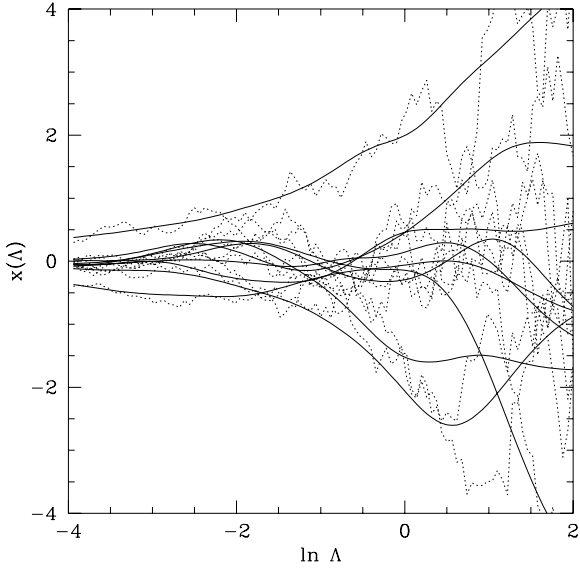,width=8cm}
}
\caption{Nine SKS and Gaussian trajectories; $n=-2$.}
\end{figure}

The main consequence of the strong correlation of trajectories in
non-SKS smoothing is that, if a trajectory is just below the absorbing
barrier, it cannot jump above it in a very small \res\ interval, as in
the SKS case. In the fixed barrier case, the PS formula {\it without
the factor of 2} is obtained at small \res. Thus, the validity of the
fudge factor 2 is limited to the very special case of SKS filtering;
any other filtering gives a number of large-mass objects that is
smaller by a factor of 2, and correspondly more small-mass objects,
thus changing the shape of the MF.

From a physical point of view, the stability of the \f\ trajectories
is a positive fact: the dynamical prediction of collapse is more
stable when \res\ varies. Gaussian filtering, in particular, has a
number of merits: it is the most stable one (it minimizes the
oscillations in the real and Fourier spaces; see also the comments in
BCEK), and it optimizes the performances of dynamical predictions
(Melott, Pellmann \& Shandarin 1994; Buchert, Melott \& Wei\ss\
1994). Unfortunately, it is mathematically much easier to work with
SKS filters, with which the \f\ trajectories are the most noisy and
least stable ones.

The main problem with Gaussian-smoothed trajectories is that, as the
filter does not cut the power spectrum in a sharp way, at a given
\res\ a trajectory contains information about the process at larger
resolutions; then, the \pot\ processes lose their Markov property. The
Langevin equation for the \pot\ processes, can be seen as an equation
of motion with {\it colored-noise} forces. All the N-point
correlations of trajectories at different resolutions are then
decisive to obtain the \pot\ PDF, which {\it does not} obey an FP
equation. Obviously, all these features are valid also for the
$\{F,\varphi\}$ system, and for the \f\ process itself, whose PDF will
not obey any FP equation.  Nonetheless, there is a motivated and
successful way, proposed by PH for linear theory ($F\propto\delta_l$),
to model Gaussian trajectories; this will be referred to as the PH
approximation.  In this approximation, the trajectories are
approximated by a step process, constant in resolution over a scale of
order \lc; different steps are assumed to be uncorrelated.  The
correlation length was chosen by PH as $\pi\mres_c\ln 2$ (see their
paper for details). The transition probability, from $\mres'$ to \res,
of such a random step process can be written as:

\beal P(F,\mres;F',\mres') &=& \delta_D(F-F') &{\rm if}& \mres/\mres'<\mres_c\\
                       &=& P(F,\mres)   &{\rm if}& \mres/\mres'\geq\mres_c\; ,
\label{eq:steppr} \eeal

\noindent from which it is possible to obtain:

\be \int_0^{F_c} \mpfu dF = \prod_i \int_0^{F_c} P_F(F,\Lambda_i) dF \; .
\label{eq:phdis} \ee

\noindent Taking a continuum limit over the logarithm of equation
(\ref{eq:phdis}), the following expression is obtained (see PH and BCEK
for details):

\bea \lefteqn{\mot = 1-P(F<F_c,\mres)}\label{eq:phcon}\label{eq:phing}
\\ &&\times\exp 
\left(\int_0^\mres \ln P(F<F_c,\mres') \frac{d\mres'}
{\pi\mres_c(\mres')\ln 2} \right) \; . \nonumber  \eea

\noindent The PH approximation has been shown, both in PH and in BCEK,
to nicely represent the true PDF, numerically calculated by simulating
a large number of Langevin trajectories. A particular feature of
Gaussian trajectories has to be noted: they explicitly depend, through
their correlation length, on the power spectrum.

\begin{figure*}
\centerline{
\psfig{figure=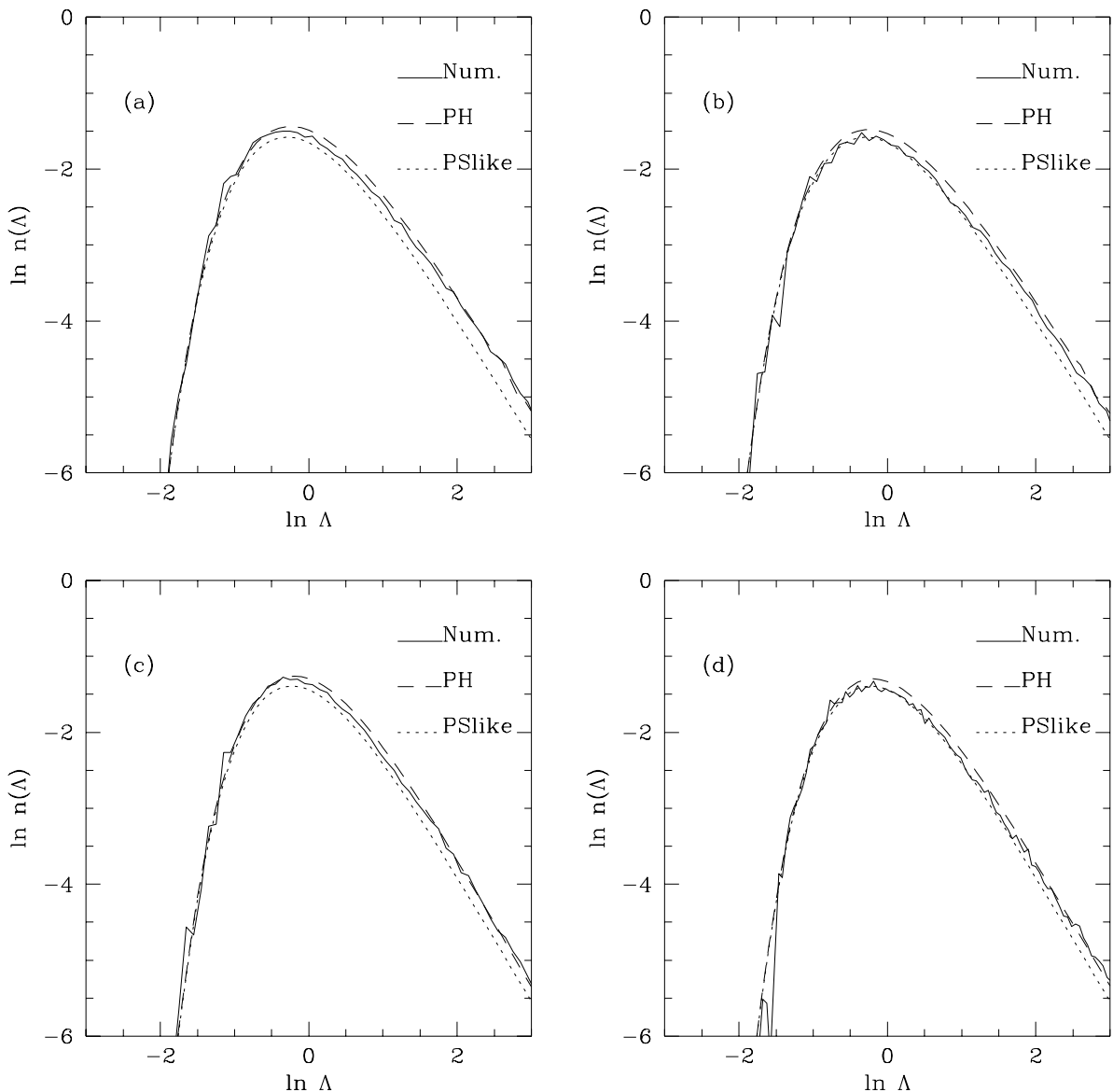,width=17cm}
}
\caption{Comparison between the PH approximation and the
numerical simulation of Langevin forces. Linear barriers have been used.
a) ELL, $n=-2$. b) ELL, $n=1$. c) 3RD, $n=-2$. d) 3RD, $n=1$.}
\end{figure*}

The PH approximation, given the uncorrelated nature of the step
process which it is based on, can be directly used to solve the moving
barrier problem with Gaussian smoothing.  Thus, expressing the
integrals in equation (\ref{eq:phing}) in the $x(F)$ variable,

\bea \lefteqn{n(\mres) = \left[\frac{\exp (-x_c^2/2\mres)}{\sqrt{2\pi\mres}} 
\left(\frac{x_c}{2\mres} - \frac{dx_c}{d\mres}
\right) + P(x<x_c,\mres) \right.} \label{eq:phnofres}\\
&&\left.\frac{\ln (P(x<x_c,\mres))}{\pi\mres_c\ln 2}\right] \exp 
\left(\int_0^\mres \ln P(x<x_c,\mres') \frac{d\mres'}
{\pi\mres_c\ln 2} \right)\; . \nonumber \eea

\noindent This expression can be compared to that obtained by means of
a PS-like procedure, as the one followed in M95:

\be \mot=1-P(x<x_c,\mres)\; ; \ee

\noindent this curve has been verified to coincide with the ELL M95
one, if the ELL barrier is used (the small-mass part is better
recovered here, as in M95 the asymptotic behavior of ellipsoidal
collapse was forced to be that of the Zel'dovich approximation). The
PH and PS-like curves coincide at large masses, and are not very
different overall (see Fig. 8). This is caused by the fact that the
integral PS-like MF is a lower limit on the true integral MF (see
BCEK), so, as only about 10 per cent of the mass has to be
redistributed, the difference between the two curves cannot be large,
especially when the correlation length is large (large $n$). This is
at variance with the SKS case, where many more objects are predicted
to form at large masses, and this makes the \mft\ curve be very
different from the PS-like one.

\begin{figure*}
\centerline{
\psfig{figure=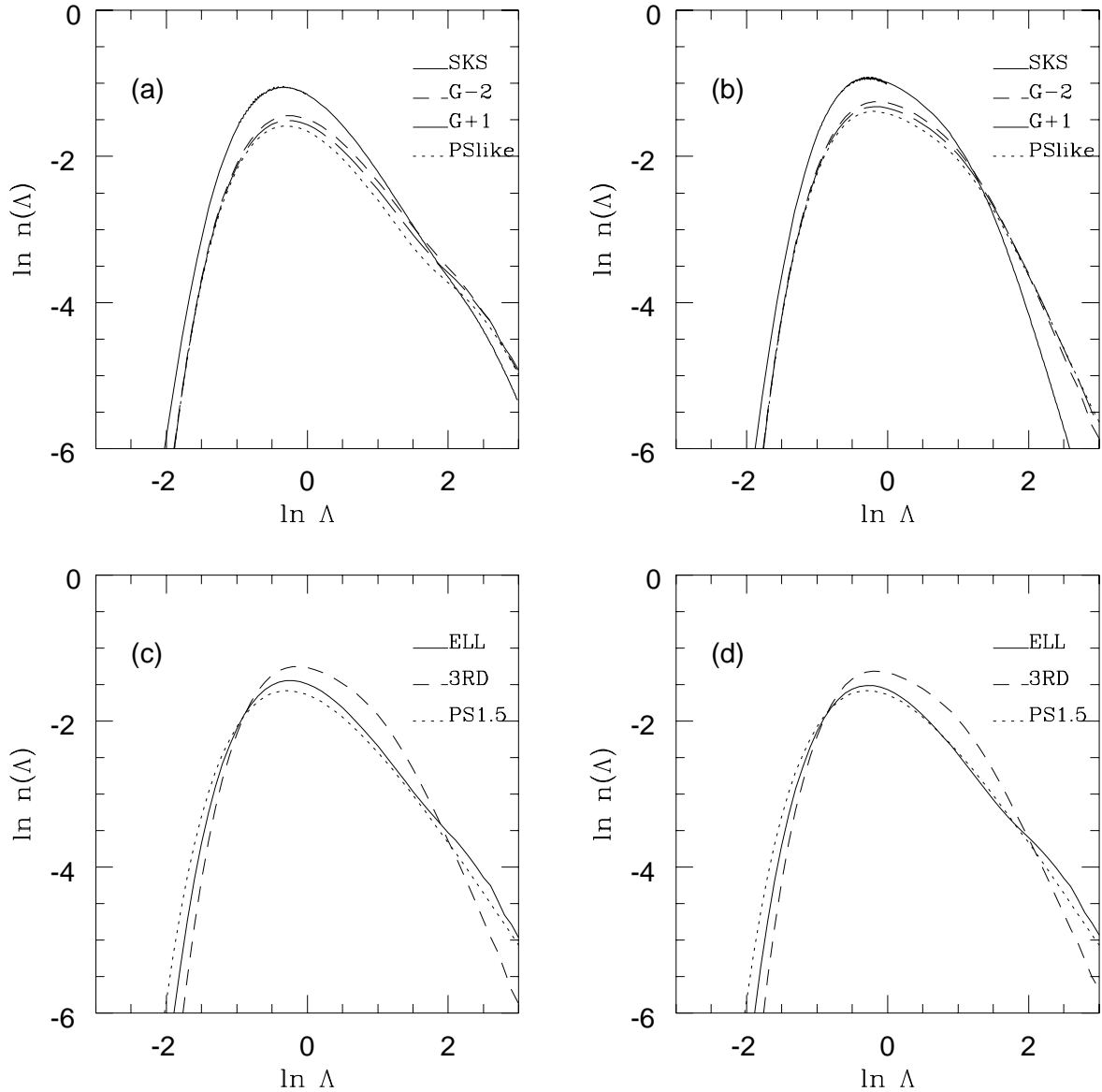,width=17cm}
}
\caption{a): ELL SKS, Gaussian ($n=-2$ and 1) and
PS-like curves. b): As a) with 3RD. c): Gaussian ELL and 3RD curves
for $n=-2$; a PS curve with $\delta_c=1.5$ is shown for
comparison. d): as c) with $n=1$.}
\end{figure*}

To test the validity of the PH approximation against direct numerical
calculations, it is possible to follow two different procedures.  The
first one, analogous to that used by Bond et al. (1991) (see also
Risken 1989) consists in simulating a large number of SKS $x$
trajectories, smoothing them by means of a Gaussian filter and finally
absorbing them with a moving $x_c(\mres)$ barrier. This procedure,
valid in the case of $\delta_l(\mres)$ trajectories, is strictly correct
(in the hypothesis that \f\ is a diffusion process, as in Section 3) if the
trajectories based on Gaussian-smoothed versions of the potential are
the same as the Gaussian-smoothed SKS \f\ trajectories; this is not
verified in general, but only if collapse time and smoothing commute.
In particular, as $F(\mres)={\cal F}[\varphi*W(\mres)]$, smoothing (in
Lagrangian space) and the functional ${\cal F}$ commute only if the
functional is linear; this is the case for spherical collapse (linear
theory with a threshold). (Another necessary assumption is the
invariance of the PDF with respect to filter shape; this is assured by
the fact that smoothing is in Lagrangian space; see Bernardeau
(1994)). A second and more correct procedure, analogous to the one
proposed in the PH paper, is to repeat the same calculations performed
in Section 3, based on the Monte Carlo realizations of paper I, with
Gaussian filtering. The first procedure has the advantage of allowing
to set both the spacing and the range in resolution as wanted,
while the second method can be used to assess the validity of the
results obtained by means of the first method.

The first procedure has been implemented as follows: 2000 random
increments have been simulated for each trajectory, in a resolution
range from $\exp(-4)$ to $\exp(4)$. The smoothed trajectories have
been computed for 100-150 resolutions; their stability makes it
unnecessary to use finer samplings. Scale-free power spectra, with
$n=-2$ and $-1$, have been used; for larger $n$ the PH and PS-like
curves are so similar that it is difficult (and not useful) to decide
which curve is better fit by the simulations.  Fig. 8 shows the
results for 50000 trajectories, compared to the PH approximations and
the PS-like curves, for linear ELL and 3RD absorbing barriers.  The
results have been rescaled to $\mres_G$, the variance of the Gaussian
process, which, for scale-free spectra, is related to the SKS one by
$\mres_G = (n+3)/2\, \Gamma((n+3)/2)\, \mres_{SKS}$ ($\Gamma$ is the
usual Gamma function).  The PH approximation accurately reproduces the
\mft\ curves, maybe slightly overestimating them around \res=1; the
asymptotic behaviors are correctly reproduced. The numerical curves
are accurate enough to prefer the PH curves with respect to the
PS-like ones, especially at large \res\ and for $n=-2$.

Figs. 9a and b compare the ELL and 3RD \mft\ curves (complete barrier)
obtained with SKS smoothing, Gaussian smoothing (PH approximation,
$n=-2$ and $-1$) and the PS-like ones.
The following things can be noted: 

\begin{enumerate}
\item the Gaussian curves are below the SKS ones by a factor of 2 in the
large- and intermediate-mass part; this surely mitigates the problem
noted above regarding the SKS curve at intermediate resolutions.
\item The small-mass slope of Gaussian curves is less steep
than the SKS ones.
\item The dependence of the \mft\ Gaussian curves on
the spectrum is modest, and only slightly affects the small-mass part.
\end{enumerate}

Figs. 9c ($n=-2$) and d ($n=1$) show the ELL and 3RD Gaussian \mft\
curves, compared to a PS curve with a $\delta_c=1.5$ value. All the
conclusions given for SKS curves, on the asymptotic behaviors,
remain valid, but this time the central part of the MF is just
slightly above the PS one. Moreover, the Gaussian curves are similar
to the PS curve with a lower $\delta_c$ value; this confirms the
findings of M95. However, any comparison with a PS curve, considered
as a fit to N-body simulations, is just qualitative, as the objects
which are described here can be different from the friends-of-friends
halos which are usually extracted from the simulations.

\begin{figure}
\centerline{
\psfig{figure=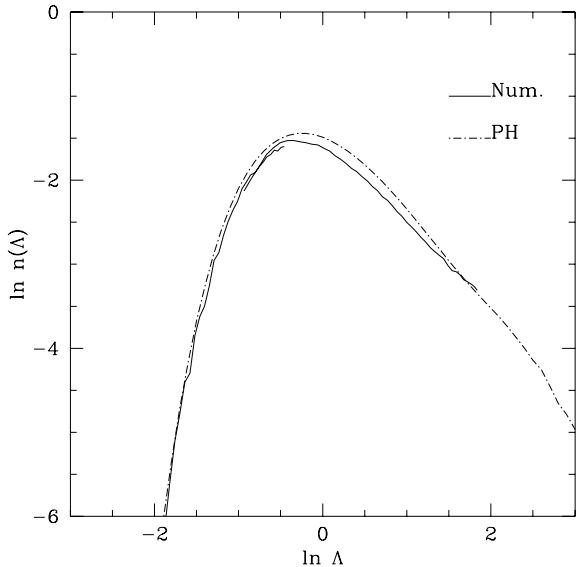,width=8cm}
}
\caption{ELL prediction: comparison of the PH
approximation with the direct simulation of the ELL collapse with
Gaussian smoothing.}
\end{figure}

The second numerical method has been implemented as follows: the
32$^3$ Monte Carlo realizations of \pot\ have been smoothed by a
hierarchy of Gaussian filters, then absorbing the obtained
trajectories with a barrier at \fc=1.  At variance with the SKS
calculations presented in Section 3, the sampling in \res\ does not
have to be very fine, as trajectories are much stable; the same
resulting \mft\ curves are much more stable than the SKS ones.  Two
sets of simulations with different normalizations have been used to
cover a significant range in \res.  Fig. 10 presents the resulting
\mft\ for the two sets of 30 realization; the ELL prediction has been
used. The PH curve is found to reproduce the numerical curve in a
satisfactory way. Its validity is then confirmed.

\section{From resolution to mass}

The quantities considered up to now, namely \ot\ and \mft, depend on
the resolution \res. To determine the MF, a relation between \res\ and
the mass $M$ is required. The collapsed medium gathers in clumps,
which are identified as structures, provided they are reasonably
separate in real space; it is necessary to determine how the mass is
distributed among those clumps. The simplest hypothesis is that a
single mass forms at every \res: $M=M(\mres)$; it is then reasonable
to assume this mass to be proportional to the cube of the smoothing
scale: $M\propto \bar{\rho} R_f^3$, as $R_f$ is the relevant
characteristic scale for the forming clump. The proportionality
constant can be obtained by connecting $M$ to the mass contained in
the smoothing filter, as in PS and in many relevant papers, or can be
left as a free parameter, as in Lacey \& Cole (1994). In the peak
approach the situation is inverted with respect to the PS ans\"atz:
the number of objects is clearly connected to the number of peaks
above a given threshold, but the mass associated to a peak, and then
the normalization of the MF itself, is not clearly determined.

With the above assumption, the MF can be easily calculated as follows:

\be n(M)dM = \frac{\bar{\rho}}{M} n(\mres(M))\left| \frac{d\mres(M)}{dM}
\right| dM\; . \label{eq:mfa} \ee

\noindent As the $\mres \rightarrow M$ transformation is a simple
functional relation, all the conclusions given above about the \mft\
quantity are valid for \mf. For scale-free power spectra, $\mres(M)$
is:

\be \mres = (M/M_*)^{-(3+n)/3}\; , \label{eq:resofm} \ee

\noindent where $M_*$ is the mass corresponding to unit variance; note
that $M_*$ is different, by a factor of $(2/\delta_c)^{3/(3+n)}$, from
the usual PS $M_*$ parameter used in the literature; since this
theory has no $\delta_c$ parameter, that factor has been omitted.
Figs. 11a ($n=-2$) and b ($n=1$) show the MFs as predicted by ELL and
3RD (with Gaussian smoothing), and a PS curve with $\delta_c=1.5$ for
comparison; because of self-similarity, the MFs are given as a
function of $M/M_*$. Spectra have been normalized by assuming unit
variance over a top-hat radius of 8$h^{-1}$ Mpc. Note how the
differences between the \mft\ curves are much less visible in the MFs,
especially for small spectral indexes; this is mainly due to the huge
dynamical range spanned by the MFs.

\begin{figure*}
\centerline{
\psfig{figure=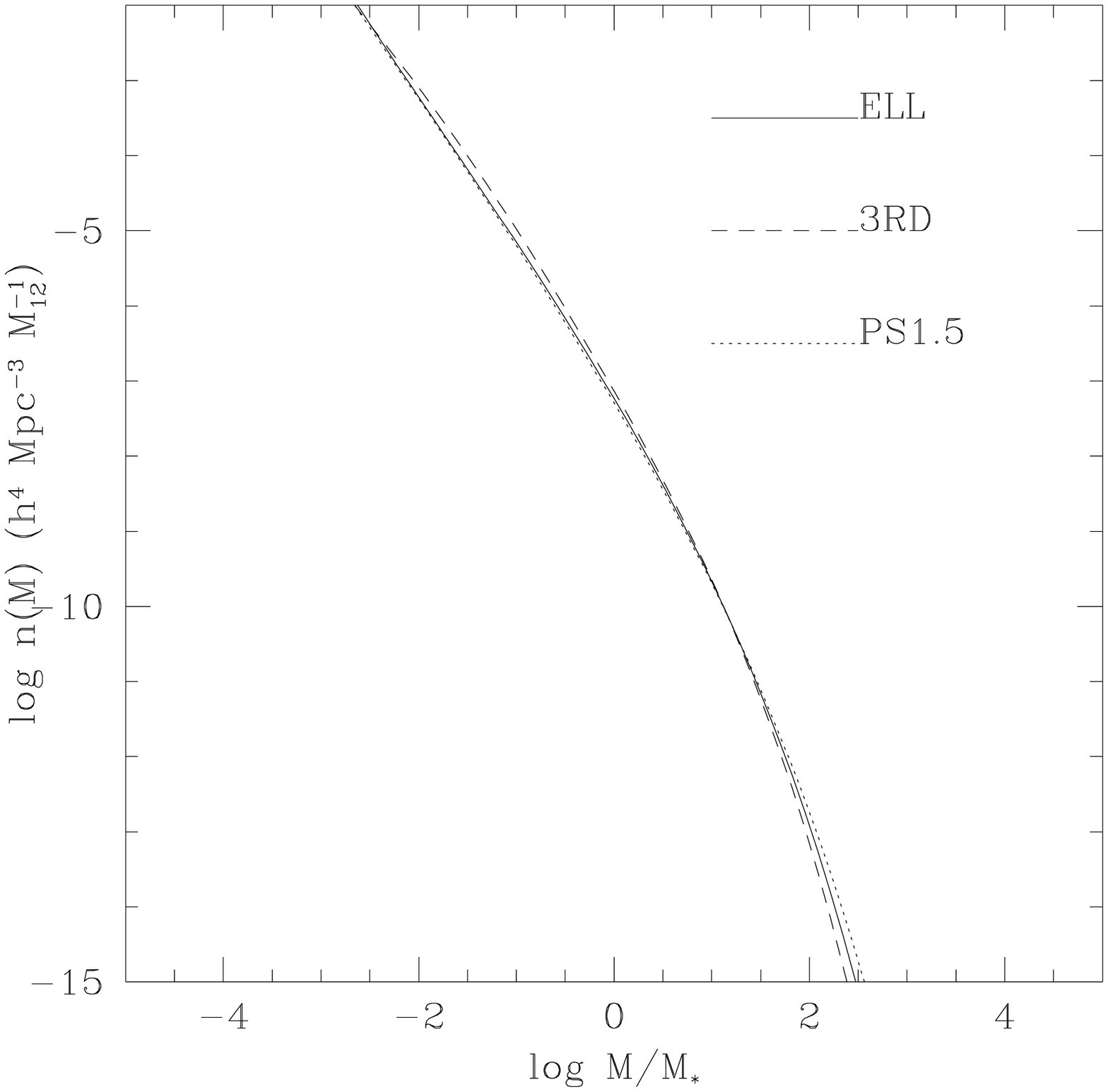,width=8cm}
\psfig{figure=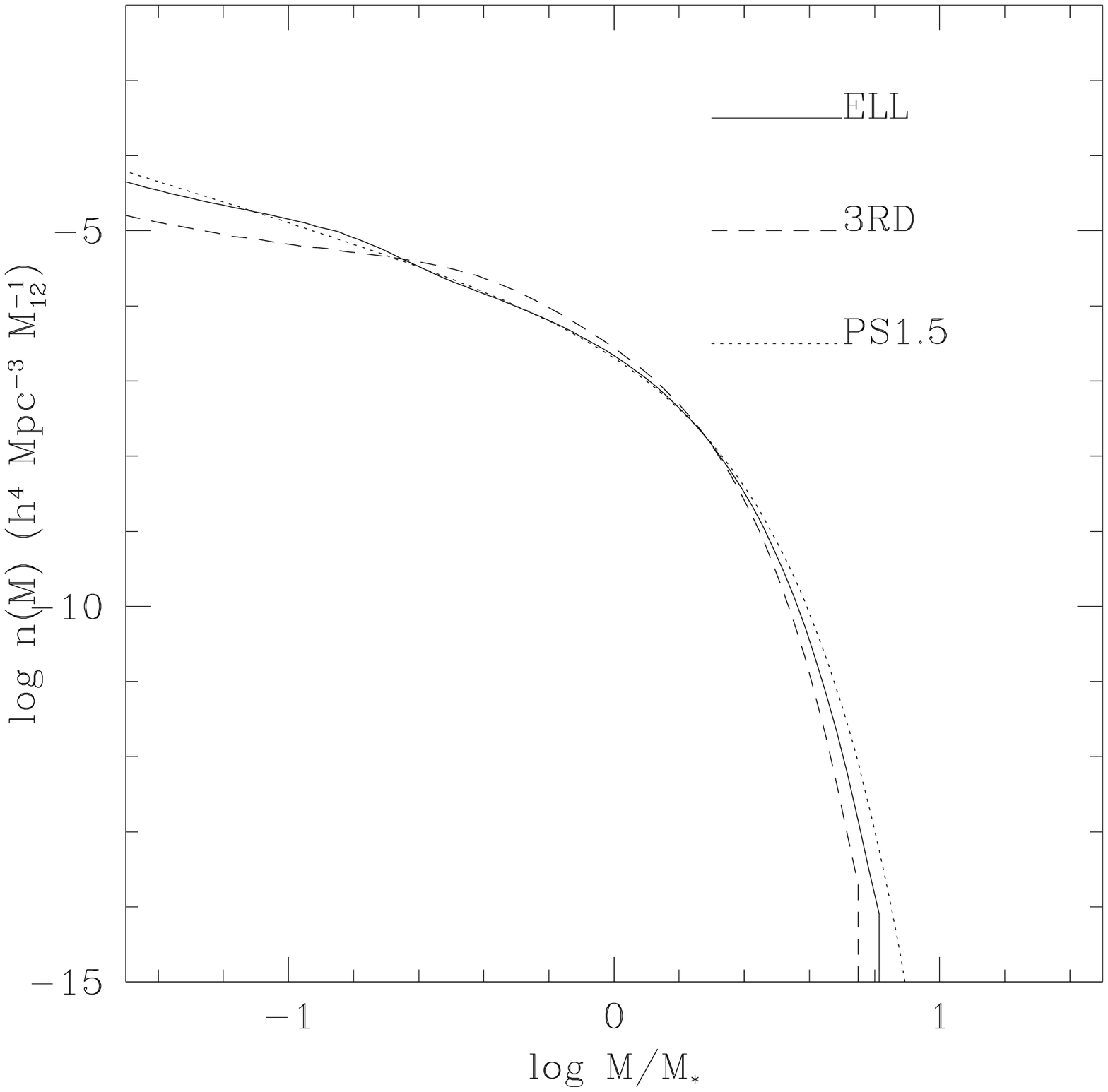,width=8cm}
}
\caption{MFs for scale-free power spectra (a):$n=-2$,
b) $n=1$), as a function of $M/M_*$: ELL and 3RD predictions with
Gaussian smoothing, and PS with $\delta_c=1.5$. $M_{12}=10^{12}M_\odot$.}
\end{figure*}

The reasonable assumption of a simple functional relation between mass
and resolution is just an approximation of what really happens.
Suppose that a whole distribution of masses is assigned to a
given resolution:

\be \mres \rightarrow p(M;\mres)\; , \label{eq:mdis} \ee

\noindent $p$ giving the probability of having a mass $M$ from a
resolution \res. Then the differential MF can be calculated as
follows:

\be n(M)dM = \frac{\bar{\rho}}{M} \left( \int_0^\infty n(\mres)
p(M,\mres) d\mres \right)\, dM \; ,\label{eq:mfdis} \ee

\noindent i.e., the \mft\ curve is convolved with the distribution of
the masses forming at a given resolution. 

The physical origin of this $p$ distribution can be explained as
follows. The real collapsed regions are related to the excursion sets
of \f\ above the \fc\ threshold, together with the points which are
predicted to collapse at smaller \res. At the large mass end, since
\f\ is roughly proportional to the density contrast (spherical
collapse is asymptotically recovered at small \f\ values), the
excursion sets are characterized by isolated, simply connected regions
in Lagrangian space, each containing a single peak (see, e.g., Adler
1981; Bardeen et al. 1986). It is then natural to assume that each
region ends up in an isolated clump. These clumps will not all have
the same mass; rather, the masses will trace a given $p$ distribution,
peaked on a given mass scale proportional to the filter mass, and with
a given width. Both mean value and dispersion will in general depend
on the power spectrum and on the shape of the filter (in this case
Gaussian filters are to be preferred, as excursion sets are much more
stable). At smaller resolutions the situation is considerably more
complex, as the topology of the excursion sets is multiply connected.
It is thus necessary to give a prescription to fragment the medium
into isolated clumps.  As a consequence, the $p$ distribution will
depend on the way the collapsed medium fragments into clumps, and its
shape will probably not be as simple as before. In conclusion, the
introduction of this $p$ distribution is not expected to dramatically
influence the MF at the large mass end; on the other hand, it is
likely to influence significantly the shape of the small-mass part.

While the punctual nature of the dynamical prediction makes it
possible to avoid taking spatial correlations explicitly into account
in determining \mft, the statistical spatial properties of the \f\
process are relevant for determining how the mass gathers into
collapsed structures.  A precise analytical determination of the
distribution $p(M;\mres)$ is prohibitive: even for Gaussian processes,
only mean values of the extension of the excursion sets can be
obtained (Adler 1981). Thus, this quantity can only be quantified
through Monte Carlo simulations and careful comparisons with extended
N-body simulations. This will be done in paper III of this series.

\section{Summary and Conclusions}

In this paper, the statistical tools, needed to determine an MF from
the PDF of the inverse collapse times given in paper I, have been
constructed. Given any inverse collapse time prediction $F=1/b_c$ (the
``time'' variable $b$ being the linear growing mode), calculated by
means of any dynamical approximation of truncated type, it has been
demonstrated numerically that, if the smoothing of the initial field
is SKS, the fraction of collapsed mass at \res, \ot, which is given by
the solution of a multi- (infinite-) dimensional FP equation, can be
calculated by extending the diffusion formalism of Bond et al. (1991)
to the \f\ process; the absorbing barrier is set at $F_c$, the inverse
of the time at which the MF is required. This procedure corresponds to
the following interpretation: when a point is collapsed according to
the prediction relative to a resolution \res, it is considered to be
collapsed at any larger resolution, even though the prediction
$F(\mres)$ does not explicitly give collapse. Note that, due to the
punctual nature of the collapse prediction, the 1-point PDF of \f\
suffices in determining the fraction of collapsed mass.

The \f\ process, assumed to be a Markov process, can be transformed to
a Gaussian Wiener process $x$ -- a random walk -- such that the \px\
distribution is a Gaussian with variance \res; this transformation has
already been found in paper I. In this way, the fixed absorbing
barrier problem for \f\ is transformed into a moving absorbing barrier
problem for $x$. This problem has been numerically solved, and a good
analytical approximation has been found in the relevant cases for
which the $F \rightarrow x$ transformation is linear; this is the case
for the large-\f\ parts of the \pf\ distributions for 3rd-order
Lagrangian perturbation theory and ellipsoidal collapse. Finally, it
has been numerically verified that the solution of the moving barrier
problem gives a correct description of the real upcrossing rate of the
\f\ process.

When the smoothing is not SKS, the $F(\mres)$ process is not a
diffusion process, as different scales are mixed by the non-sharp
truncation of the power spectrum. In this case, the \f\ process is
much more stable, so that a good approximation of it can be obtained
by considering it constant over its correlation length. Thus, the
useful approximation developed by Peacock \& Heavens (1990) can be
used.  The PH approximation adequately reproduces the numerical
simulations (performed by means of two different methods) of Langevin
trajectories absorbed by a moving barrier.  The advantages of dealing
with Gaussian smoothing are its physical meaning (Gaussian smoothing
is usually recommended when using truncated dynamical approximations;
see paper I) and its stability with respect to \res. The obvious
disadvantage is that it is much more complicated to solve the
absorbing-barrier problem.

To determine the number of collapsed objects from the fraction of
collapsed mass at a given \res, information is needed about how the
collapsed points gather together in extended structures. This is where
the spatial correlations of the \f\ process are decisive. A first
approximation can be given by assuming that a mass equal to that
contained in the smoothing filter is formed.  More realistically, a
whole distribution $p(M;\mres)$ of masses has to be associated to
every resolution; this distribution, which will depend on the power
spectrum, is likely to affect the small-mass part of the MF considerably. A
purely analytical determination of the $p$ distribution is problematical
even in the case of Gaussian processes; its study is postponed to the
forthcoming paper III.

The following final conclusions can be drawn on the MF:

\begin{enumerate}
\item A larger number of large-mass objects is expected to form
with respect to the simple PS prediction; this agrees with the
conclusions of M95.
\item The large-mass part of the MF is robust with respect to the
dynamical prediction: different reasonable dynamical predictions give
similar results. The ELL prediction tends to give more objects than
the 3RD one in the large-mass tail; this is due to the fact that
3rd-order Lagrangian theory slightly underestimates spherical
collapse; thus the ELL prediction is considered more believable in
that range.
\item Explicit, reasonably accurate, analytical solutions are
given for the large-mass parts of the SKS and Gaussian MFs.
\item When Gaussian smoothing is used, the resulting MFs give fewer
objects by roughly a factor of 2 in the large-mass part than the SKS
one.  This makes the Gaussian curves very similar to the PS one with a
$\delta_c\simeq1.5$ parameter.
\item The Gaussian MFs are preferred to the SKS ones because
Gaussian smoothing optimizes the dynamical predictions, stabilizes the
trajectories, and lowers the peak of the MF. 
\end{enumerate}

Note that the dynamical predictions analyzed here, particularly the
3rd-order Lagrangian one, tend to produce more intermediate-mass
objects, at the expense of the small-mass ones, somehow introducing a
second characteristic scale in the MF.

The small-mass part of the MF is not considered a robust prediction of
the theory, for at least three reasons:

\begin{enumerate}
\item The definition of collapse, given in paper I, which is
based on the concept of orbit crossing, is not expected to reproduce
common small-mass structures like virialized halos. OC regions rather
represent those large-scale collapsed environments in which the
virialized halos are embedded. 
\item All the dynamical predictions given in paper I are
considered good as long as the inverse collapse time is not
small. Thus, the small-mass part of the MF is based on non-robust
dynamical predictions.
\item The $p$ distribution of the forming masses, at a given
resolution, is likely to dramatically affect the small-mass part of
the MF. 
\end{enumerate}

\section*{Acknowledgments}

I wish to thank Alfonso Cavaliere for a number of discussions.  The
help of Gianfausto Dell'Antonio, Davide Gabrielli and Luca Sbano has
been precious in dealing with stochastic processes. I also thank Paolo
Catelan and Cristiano Porciani for useful discussions and an anonymous
referee for his constructive criticisms. This work has been partially
supported by the Italian Research Council (CNR-GNA) and by the
Ministry of University and of Scientific and Technological Research
(MURST).


\appendix


\section{Alternative Formulation of the FP equation}

Equation (\ref{eq:fpx}), with a moving absorbing barrier, can be
expressed in a form which includes the barrier condition in the drift
and diffusion coefficients; this can help in finding analytical
solutions.  Let $\xi(t)$ be a Wiener process, and let $x$ be the value
it takes at time $t$ (the process was called $x(t)$ in the text; $t$
corresponds to \res). The $P_x(x,t)$ PDF obeys an evolution equation
of the kind:

\be \frac{\partial P}{\partial t} = \sum_{n=1}^\infty \left( -\frac{
\partial}{\partial x} \right)^n\, D^{(n)}(x,t)\, P \; .\label{eq:kmfp} \ee

Drift and diffusion coefficients are the first two of the series; all
the other coefficients vanish for continuous Markov processes. The
coefficients can be found by means of the Kramers-Moyal expansion
(Risken 1989):

\bea \lefteqn{D^{(n)}(x,t) = \frac{1}{n!} \lim_{\Delta t\rightarrow 0} 
\frac{1}{\Delta t} \left. \langle (\xi(t+\Delta t)-\xi(t))^n \rangle 
\right|_{\xi(t)=x}}\label{eq:kmexp}\\ &&=\frac{1}{n!} \lim_{\Delta 
t\rightarrow 0} \frac{1}{\Delta t}\int(x-x')^n P(x,t+\Delta t|x',t) dx\; . 
\nonumber \eea

These coefficients can be easily calculated by integrating the Gaussian
transition probability:

\be P(x,t+\Delta t|x',t) = \exp(-(x-x')^2/2\Delta t)/\sqrt{2\pi\Delta t}
\label{eq:gtrans} \ee

\noindent in the relevant integration range, i.e. from $-\infty$ to
$x_c(t)$.  The result of this integration is:

\bea D^{(1)}(x,t) &=& -\delta_D(x_c(t)-x) \nonumber\\ 
     D^{(2)}(x,t) &=& \theta(x_c(t)-x)/2\label{eq:kmcoef} \\
     D^{(n)}(x,t) &=& 0 \ \ \ \ \ \ \ \ n>2\; .\nonumber  \eea

\noindent The interpretation of this result is simple: the upcrossing
of the barrier is contrasted by an infinite discontinuous drift, while
diffusion is switched off beyond the barrier. With some algebraic
manipulation, the equation can be written as:

\be \frac{\partial P}{\partial t} = -\frac{1}{2} \frac{\delta_D(x-xc(t))}
{x-xc(t)} P + \frac{1}{2}\theta(xc(t)-x)\frac{\partial^2 P}{\partial x^2}\; .
\label{eq:nicefp} \ee

The first term of the right hand side,linear in $P$, can be
interpreted as a branching term which kills the trajectories
upcrossing the barrier (see Cavaliere et al. 1996).  The solution of
this equation can be written in terms of a functional integral; it is
easy to show that (see Risken 1989, Section 4.4.2):

\bea \lefteqn{\mpxu = \lim_{N\rightarrow\infty} \int_{-\infty}^{x_c(\mres)}
\!\!\!\!\!\! dx_{N-1} \ldots \int_{-\infty}^{x_c(\mres_{i+1})} \!\!\!\!\!\!
dx_i \cdots\int_{-\infty}^{x_c( \mres_1)} \!\!\!\!\!\!dx_0}\nonumber\\ 
&&\prod_{i=0}^{N-1} 
\frac{1}{\sqrt{2\pi d\mres_i}} \exp \left[ -\frac{(x_{i+1}-x_i)^2}{2d\mres}
\right]\; . \label{eq:path} \eea

\noindent Unfortunately, the variable upper limits make this integral
very hard to solve.

\bsp

\label{lastpage}


\begin{thebibliography}{99}
\bibitem{b} Adler R.J., 1981, The Geometry of Random Fields. Wiley, New York
\bibitem{b} Appel L., Jones B.J.T., 1990, MNRAS, 245, 522
\bibitem{b} Arnold L., 1973, Stochastic Differential Equations. Wiley, New York
\bibitem{b} Bardeen J.M., Bond J.R., Kaiser N., Szalay A.S., 1986, ApJ, 304, 15
\bibitem{b} Bernardeau F., 1994, A\&A, 291, 697
\bibitem{b} Blanchard A., Valls-Gabaud D., \& Mamon G.A., 1992, A\&A, 264, 365
\bibitem{b} Bond J.R., Cole S., Efstathiou G., Kaiser N., 1991, ApJ, 379, 440
(BCEK)
\bibitem{b} Bond J.R., Myers S.T., 1996, ApJS, 103, 1
\bibitem{b} Bouchet F.R., Colombi S., Hivon E., Juszkiewicz R., 1995, A\&A, 
296, 575
\bibitem{b} Bower R.G., 1991, MNRAS, 248, 332
\bibitem{b} Buchert T., 1994, MNRAS, 267, 811
\bibitem{b} Buchert T., Melott A.L., Wei\ss\  A.G., 1994, A\&A, 288, 349
\bibitem{b} Catelan P., 1995, MNRAS, 276, 115
\bibitem{b} Catelan P., Lucchin F., Matarrese S., 1988, Phys.Rev.Lett, 61, 267
\bibitem{b} Cavaliere A., Colafrancesco S., Scaramella R., 1991, ApJ, 380, 15
\bibitem{b} Cavaliere A., Menci N., Tozzi P., 1994, in Seitter W.C., ed., 
Cosmological Aspects of X-ray Clusters of Galaxies.  Kluwer Ac. Pub., Dordrecht
\bibitem{b} Cavaliere A., Menci N., Tozzi P., 1996, ApJ, 464, 44
\bibitem{b} Chandrasekhar S., 1943, Rev. Mod. Phys., 15, 2
\bibitem{b} Colafrancesco S., Lucchin F., Matarrese S., 1989, ApJ, 345, 3
\bibitem{b} Cole S., Kaiser N., 1989, MNRAS, 237, 1127
\bibitem{b} Epstein R.I., 1983, MNRAS, 205, 207
\bibitem{b} Katz N., Quinn T., Gelb J.M., 1993, MNRAS, 265, 689 
\bibitem{b} Lacey C., Cole S., 1993, MNRAS, 262, 627
\bibitem{b} Lacey C., Cole S., 1994, MNRAS, 271, 676
\bibitem{b} Lilje P.B., 1992, ApJ, 386, L33
\bibitem{b} Lucchin F., 1989, in Flin P., Duerbeck H.W., eds., Morphological 
Cosmology. Springer Verlag, Berlin 
\bibitem{b} Lucchin F., Matarrese S., 1988, ApJ, 330, 535
\bibitem{b} Manrique A., Salvador-Sol\'e E., 1995, ApJ, 453, 6
\bibitem{b} Manrique A., Salvador-Sol\'e E., 1996, ApJ, 467, 504
\bibitem{b} Melott A.L., Pellman T., Shandarin S.F., 1994, MNRAS, 269, 626
\bibitem{b} Monaco P., 1995, ApJ, 447, 23 (M95)
\bibitem{b} Monaco P., 1996, MNRAS, in press (paper I)
\bibitem{b} Peacock J.A., Heavens A.F., 1990, MNRAS, 243, 133 (PH)
\bibitem{b} Porciani C., Ferrini F., Lucchin F., Matarrese S., 1996, MNRAS,
281, 311
\bibitem{b} Press W.H., Flannery B.P., Teukolsky S.A., Vetterling W.T., 1992,
Numerical Recipes in Fortran. Cambridge University Press, Cambridge
\bibitem{b} Press W.H., Schechter P., 1974, ApJ, 187, 425 (PS)
\bibitem{b} Risken H., 1989, The Fokker-Planck Equation. Springer Verlag, 
Berlin  
\bibitem{b} Rodriguez D.D.C., Thomas P.A., 1996, MNRAS, 282, 631
\bibitem{b} Ryden B.S., 1988, ApJ, 333, 78
\bibitem{b} Shandarin S.F., Zel'dovich Ya.B., 1989, Rev. Mod. Phys., 61, 185
\bibitem{b} Silk J., White S.D., 1978, ApJ, 223, L59
\bibitem{b} van de Weygaert R., Babul A., 1994, ApJ, 425, L59
\bibitem{b} Vergassola M., Dubrulle B., Frisch U., Noullez A., 1994, A\&A, 
289, 325
\bibitem{b} Yano T., Nagashima M., Gouda N., 1996, ApJ, 466, 1
\bibitem{b} Zel'dovich Ya.B., 1970, Astrofizika, 6, 319 (transl.: 1973, 
Astrophysics, 6, 164)
\end{thebibliography}
\end{document}